%% file: root.tex
\documentclass[5p,times]{elsarticle}  % Comment this line out if you need a4paper

\usepackage{siunitx}
\usepackage{microtype}
\usepackage{amsmath} % assume amsmath package installed
\usepackage{amssymb}  % assumes amsmath package installed

\usepackage{amsthm}
\usepackage{booktabs}
\usepackage{algorithm} 
\usepackage{algpseudocode} 
\usepackage{enumitem}
\newtheorem{lemma}{Lemma}[section]
\usepackage{subcaption}
\usepackage{tikz}
\usetikzlibrary{shapes,arrows,calc,fit}
\tikzset{
        block/.style = {draw, rectangle,
            minimum height=1cm,
            minimum width=2cm},
        input/.style = {coordinate,node distance=1cm},
        output/.style = {coordinate,node distance=4cm},
        arrow/.style={draw, -latex,node distance=2cm},
        pinstyle/.style = {pin edge={latex-, black,node distance=2cm}},
        sum/.style = {draw, circle, node distance=1cm},
    }
\usepackage{pgfplots} 
\pgfplotsset{compat=newest} 
\pgfplotsset{plot coordinates/math parser=false} 
\newlength\figureheight% 
\newlength\figurewidth% 
\pgfplotsset{
    every axis plot post/.style={
        line join=round
    }
}
\usepackage{algorithm} 
\usepackage{algpseudocode} 
\newlength\defcolwidth%

\input{tikz_set}

\newtheorem{theorem}{Theorem}

\journal{journal}
\usepackage{eso-pic}

\AddToShipoutPictureBG*{%
	\AtPageUpperLeft{%
		\setlength\unitlength{1in}%
		\hspace*{\dimexpr0.5\paperwidth\relax}%% change \dimexpr0.5\paperwidth\relax appropriately
		\makebox(0,-1)[c]{\begin{tabular}{c c}
				Max van Meer, Compensating Hysteresis and Mechanical Misalignment in Piezo-Stepper Actuators, \\
				Submitted to journal March 14, 2025,\\Uploaded to ArXiv \today
		\end{tabular}}
}}

\begin{document}
\begin{frontmatter}
    \title{Compensating Hysteresis and Mechanical Misalignment in Piezo-Stepper Actuators\tnoteref{t1}}
    \tnotetext[t1]{This work is part of the research program VIDI with project number 15698, which is (partly) financed by the Netherlands Organisation for Scientific Research (NWO). In addition, this research has received funding from the ECSEL Joint Undertaking under grant agreement 101007311 (IMOCO4.E). The Joint Undertaking receives support from the European Union's Horizon 2020 research and innovation program.}

    \affiliation[tue]{organization={Department of Mechanical Engineering, Control Systems Technology},%Department and Organization
            addressline={Eindhoven University of Technology}, 
            city=Eindhoven,
            address={PO Box 513, 5600MB,},
            country={the Netherlands}}
\affiliation[tfs]{organization={Thermo Fisher Scientific},
            city=Eindhoven,
            country={the Netherlands}}     
\affiliation[tno]{organization={Department of Optomechatronics},%Department and Organization
            addressline={TNO}, 
            city=Delft,
            country={the Netherlands}}            
\affiliation[tud]{organization={Delft Center for Systems and Control},%Department and Organization
            addressline={Delft University of Technology}, 
            city=Delft,
            country={the Netherlands}}        
% Define authors with their respective affiliation markers
\author[tue]{Max van Meer\corref{cor1}}\ead{m.v.meer@tue.nl} % Multiple affiliations example
\cortext[cor1]{Corresponding author}
% \address[tue]{Eindhoven University of Technology, Eindhoven, the Netherlands}
\author[tue]{Tim van Meijel}
\author[tfs]{Emile van Halsema}
\author[tfs]{Edwin Verschueren}
\author[tue,tno]{Gert Witvoet} % Multiple affiliations example
\author[tue,tud]{Tom Oomen}

\begin{abstract}
    Piezo-stepper actuators enable accurate positioning through the sequential contraction and expansion of piezoelectric elements, generating a walking motion. The aim of this paper is to reduce velocity ripples caused by parasitic effects, due to hysteresis in the piezoelectric material and mechanical misalignments, through suitable feedforward control. The presented approach involves the integration of a rate-dependent hysteresis model with a position-dependent feedforward learning scheme to compensate for these effects. Experimental results show that this approach leads to a significant reduction in the velocity ripples, even when the target velocity is changed. These results enable the use of piezo-stepper actuators in applications requiring high positioning accuracy and stiffness over a long stroke, without requiring expensive position sensors for high-gain feedback.
\end{abstract}
\end{frontmatter}

\section{Introduction}
Piezo-stepper actuators offer a promising solution for nano-scale positioning by combining high stiffness, a long stroke, and precise motion control. High stiffness minimizes deformation under load, while a long stroke extends their applicability to tasks requiring significant range of motion~\cite{Fleming2014,Li2019a,Kanchan2023}. These features make them well-suited for applications such as scanning tunneling microscopy (STM) and electron microscopy (EM)%, and vibration isolation
~\cite{DenHeijer2014a, Mohith2021, Strijbosch2021}. % Removed Lu2020

\begin{figure}
    \centering
    \includegraphics[width=\linewidth]{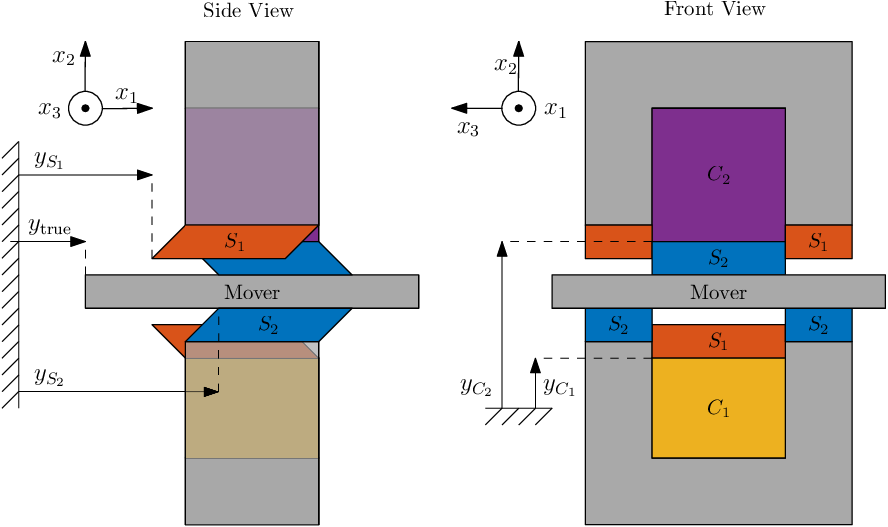}
    \caption{Schematic of a piezo-stepper actuator. The clamps ($C_1$, $C_2$) press the shear elements ($S_1$, $S_2$) onto the mover. When a shear element $S_e$ is in contact with the mover, it expands or contracts laterally to push or pull the mover in the $x_1$ direction.}\label{fig:piezo_scheme}

    \end{figure}

\begin{figure}
        \centering
        \includegraphics[width=\linewidth]{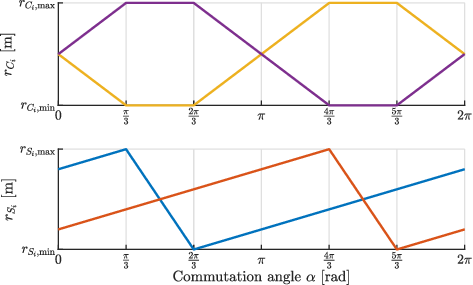}
        \caption{Reference displacements for piezo elements to achieve a stepping motion. The clamps (\protect\yelline,\protect\purline) press the shears (\protect\blueline,\protect\redline) onto the mover one by one, and the shears drag the mover along in the lateral direction. }\label{fig:waveforms}
        \end{figure}

Multiple piezo-stepper designs are available for different purposes. For EM and STM, the design relies on an ingenious stacking of piezoelectric elements, see Figure~\ref{fig:piezo_scheme}. Clamp elements expand vertically to press shear elements onto a central mover, while the shear elements expand laterally to displace the mover. When the shears reach the end of their stroke, the clamps disengage, allowing a second set of clamps and shears to take over, generating a walking motion over a large stroke. Figure~\ref{fig:waveforms} illustrates the idealized reference displacements for the individual piezo elements during this process.

\begin{figure}
                \centering
                \begin{subfigure}{0.48\linewidth}
                    \centering
            \includegraphics[width=\linewidth]{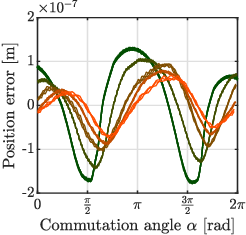}
                    \caption{Forwards stepping motion}
                \end{subfigure}
                \hfill
                \begin{subfigure}{0.48\linewidth}
                    \centering
                    \includegraphics[width=\linewidth]{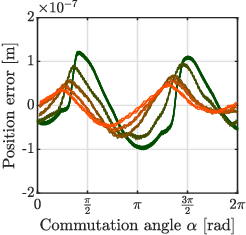}
                    \caption{Backwards stepping motion}
                \end{subfigure}
                \caption{Position error of the mover against the commutation angle $\alpha$, using voltage waveforms that scale with the references in Figure~\ref{fig:waveforms}, for a range of constant drive frequencies between 0.4 Hz (\protect\dgreenline) and 100 Hz (\protect\dredline), showing three steps per frequency. The data shows a direction-dependent error that remains consistent across steps, suggesting it is caused by $\alpha$-domain disturbances. Variations across drive frequencies are attributed to a combination of history-dependent hysteresis effects and lowpass effects of the capacitive position sensor.}\label{fig:alpha-dependent}
                        \end{figure}

The stacking of piezo elements introduces a number of parasitic effects that must be compensated in high-accuracy applications: hysteresis and mechanical misalignments. Hysteresis, an inherent property of piezoelectric materials, leads to a history-dependent relationship between the applied voltage and the resulting displacement. Mechanical misalignments, on the other hand, arise due to manufacturing and assembly tolerances, causing the piezo elements to expand or retract in unintended directions. When left unaddressed, these misalignments result in velocity ripples of the mover (see Figure~\ref{fig:alpha-dependent}). Both effects significantly impact positioning performance and must be addressed to be applicable in high-accuracy applications.

Existing solutions to mechanical misalignments have achieved notable improvements in performance. Position feedback control suppresses velocity ripples~\cite{Ruan2024}, improving accuracy while retaining flexibility. Model-based optimization of voltage waveforms~\cite{merry2011} further improves stepping performance. Data-driven feedforward methods, such as Iterative Learning Control (ILC) \cite{Li2005}, refine reference displacements using measured position data, achieving nearly constant velocity~\cite{Aarnoudse2023,Strijbosch2019}. Despite these achievements, existing solutions introduce critical limitations. Feedback control requires high bandwidth over an additional, possibly expensive, position sensor to be effective, amplifying noise and limiting suitability for nano-scale positioning~\cite{franklinFeedback}. Model-based approaches require costly parameter identification, and ILC-based methods compensate for hysteresis when not explicitly addressed, resulting in history-dependent references that reduce task flexibility.

Hysteresis compensation methods have also shown progress in improving positioning accuracy. Operator-based models, such as the Prandtl-Ishlinskii and Preisach models, capture the nonlinear, history-dependent behavior of hysteresis and are widely used in feedforward control~\cite{Strijbosch2023, Ge1995, Li2015, Kuhnen2003}. Differential-based models, including Bouc-Wen, Dahl, and Duhem models~\cite{Hassani2014}, and direct inverse hysteresis modeling~\cite{Yang2015, Qin2013}, have also proven effective. Piezo displacement measurements, required for hysteresis modeling, can even successfully be estimated from measurements of the current, reducing component costs~\cite{Furutani1998,Ronkanen2002}. Despite their strengths, these methods face challenges when applied to feedforward control. Operator-based models are computationally expensive, complicating real-time control, while differential-based models are difficult to identify reliably~\cite{Gan2019}. Direct inverse modeling, although avoiding inversion steps, complicates validation against experimental data. Data-driven feedforward approaches often rely on piezo element displacement measurements, which are impractical in applications with limited space or cost constraints, and while piezo displacement estimation from measurements of currents has proven useful for feedback control, literature on its application to hysteresis feedforward is sparse. %Therefore, computational efficiency and the use of alternative measurements, such as current, are important considerations for hysteresis feedforward. 
% Kang Liang?
% Dermich?

Although significant progress has been made in improving the performance of piezo-stepper actuators, no existing method compensates for both mechanical misalignments and hysteresis without amplifying noise, requiring accurate models, limiting flexibility, or demanding high computational power. The aim of this paper is to develop a unified framework combining rate-dependent hysteresis compensation with data-driven learning for mechanical misalignments, relying solely on measurements of the piezo currents and mover position. The framework is designed for real-time hardware to achieve high positioning accuracy in industrial applications. While the developed approach is demonstrated on a specific piezo-stepper actuator, the framework is general and can be applied to other piezo-stepper actuators with similar challenges.

The contributions of this paper are therefore as follows:
\begin{enumerate}[label=C\arabic*] 
    \item A rate-dependent hysteresis compensation method is developed that relies on measurements of the currents, eliminating the need for direct displacement sensors. The model uses few parameters and supports real-time implementation.
    \item A data-driven approach is developed to address mechanical misalignments, incorporating iterative learning control to refine reference displacements. The learned compensation function is applicable to arbitrary reference motions.
    \item These two methods are integrated into a unified feedforward control framework, ensuring high positioning accuracy across varying reference motions without requiring separate tuning per task.
    \item The framework is experimentally validated on a piezo-stepper actuator, demonstrating its effectiveness in industrially relevant scenarios.
\end{enumerate}

This paper is structured as follows. First, Section~\ref{sec:problem} formalizes the problem. Section~\ref{sec:hysteresis} presents the rate-dependent hysteresis compensation method. Subsequently, Section~\ref{sec:misalignments} describes the compensation  of mechanical misalignments and integrates both solutions into a unified framework. Section~\ref{sec:experimental} presents experimental results, and Section~\ref{sec:conclusion} concludes the paper.

\section{Problem Formulation}\label{sec:problem}
This section describes the problem of accurate control of piezo-stepper actuators with task flexibility, starting with a description of the experimental setup that serves as a foundation for the problems addressed in this paper. 

\subsection{Experimental setup}\label{sec:setup}
\begin{figure}
    \centering
    \includegraphics[width=0.9\linewidth]{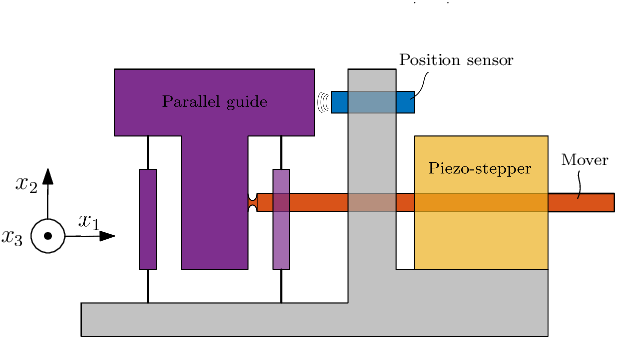}
    \caption{Schematic overview of the experimental setup. The lateral displacement $y$ of the mover is measured using a capacitive sensor via a parallel guide.}\label{fig:piezo_setup}
    \end{figure}

The experimental setup, shown schematically in Figure~\ref{fig:piezo_setup}, consists of a piezo-stepper actuator provided by Thermo Fisher Scientific. The actuator achieves a range of \SI{500}{\micro\meter} by stepping in intervals of up to \SI{3}{\micro\meter}. The displacement $y$ of the mover is measured using a capacitive sensor with a sampling rate of 10 kHz. 

The actuator operates via two groups of piezo elements, each containing a set of shear elements ($S_1, S_2$) and a clamp element ($C_1, C_2$). As shown in Figure~\ref{fig:piezo_scheme}, the shears expand and contract laterally to move the central mover, while the clamps alternately press the shears onto the mover to enable a walking motion. The current through the piezo elements is recorded and available for offline estimation of the piezo displacements, as direct measurements of individual displacements are unavailable. This setup provides the platform for addressing the challenges of hysteresis and misalignments in piezo-stepper actuators.

\subsection{Piezo element dynamics}
% Explain just enough to lay the groundwork for Fig. 4. That includes:
% - M, at a basic level
% - \kappa mover position
The rate of displacement $\dot{y}_e(t)$ of a piezo element \(e \in \Omega\) with $\Omega:=\{S_1,S_2,C_1,C_2\}$, neglecting creep, is described by~\cite{Strijbosch2023}:
\begin{equation}\label{eq:hysteresis}
    \dot{y}_e(t) = M_e(\dot{u}_e(t), u_{e,a}(t)) \dot{u}_e(t),
\end{equation}
where \(u_e(t)\) is the voltage applied to element \(e\), \(M_e\) is a hysteresis function, and \(u_{e,a}(t)\) is the voltage absement, defined formally later. The absement captures voltage history in a single parameter, which has proven effective for modeling piezoelectric hysteresis~\cite{Strijbosch2023}. 
The motion of the mover depends on the displacement of all piezo elements, as defined by:
\begin{equation}\label{eq:mover}
\begin{aligned}
    \dot{y}_{\text{true}}(t) &= \dot{y}^{\circ}(t) + \dot{d}(t), \\
    \dot{y}^{\circ}(t)&:=\kappa\big(\dot{y}_{S_1}(t), \dot{y}_{S_2}(t), \dot{y}_{C_1}(t), \dot{y}_{C_2}(t)\big)\\
    \kappa(\cdot) &= 
    \begin{cases}
        \dot{y}_{S_1} & \text{if } y_{C_1} \geq y_{C_1,\text{c}} \text{ and } y_{C_2} < y_{C_2,\text{c}}, \\
        \dot{y}_{S_2} & \text{if } y_{C_2} \geq y_{C_2,\text{c}} \text{ and } y_{C_1} < y_{C_1,\text{c}}, \\
        \frac{1}{2} (\dot{y}_{S_1}+\dot{y}_{S_2}) & \text{if } y_{C_1} \geq y_{C_1,\text{c}} \text{ and } y_{C_2} \geq y_{C_2,\text{c}}, \\
        0 & \text{otherwise}.
    \end{cases}
\end{aligned}
\end{equation}
Here, \(y_{C_i,\text{c}}\) are constants representing when clamps bring the shears into contact with the mover, and \(\dot{d}(t)\) accounts for velocity disturbances such as slip and mechanical misalignments. The measured mover position, sampled at intervals \(T_s = 1/F_s\), is expressed as:
\begin{equation}\label{eq:drag2}
    y(t_k) = y(t_0) + G(q) \dot{y}_{\text{true}}(t_k),
\end{equation}
where $q$ is the forward-shift operator in discrete-time such that $q t_k = t_{k+1}$ and $k$ is the sample number. The sensor $G(q)$ consists of a discrete-time integrator, a delay, and lowpass dynamics with a cutoff frequency around 100 Hz. Finally, it is assumed that $y(t_0)=\dot{y}(t_0)=0$. The next section describes the problem of compensating the first parasitic effect, hysteresis.

\subsection{Hysteresis compensation}\label{sec:traditional}
\begin{figure}
    \centering
    \includegraphics[width=\linewidth]{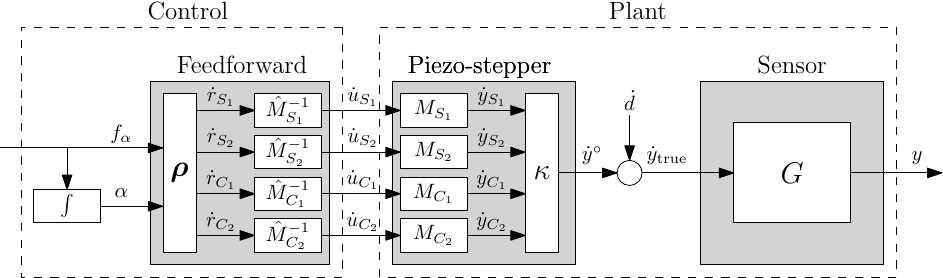}
    \caption{Schematic depiction of feedforward control of a piezo-stepper actuator. The feedforward controller yields voltages based on a commutation angle $\alpha$. When applied to the piezo elements $e$, their individual motions $\dot{y}_e$ cause a displacement of the mover via kinematics $\kappa$. % The mover position is measured by a sensor with frequency response $G(q)$. 
    Mechanical misalignments in the piezo elements lead to a velocity disturbance $\dot{d}$.}\label{fig:control_scheme}
    \end{figure}

Piezo-stepper actuators exhibit hysteresis, causing a history-dependent relationship between voltage and displacement, see~\eqref{eq:hysteresis}. To compensate for this, a model $\hat{M}_e$ approximates the true hysteresis function $M_e$ for every element and is inverted in a feedforward control law. When $\hat{M}_e\approx M_e$, the required voltages are given by
\begin{equation}\label{eq:inv_hysteresis}
    \dot{u}_e = \hat{M}_e^{-1} \dot{r}_e,
\end{equation}
resulting in accurate tracking $\dot{y}_e\approx\dot{r}_e$ of the reference displacements. These reference displacements $\dot{r}_e$ are defined by a waveform function $\boldsymbol{\rho}$ as
    %  \begin{equation*}
    %     \boldsymbol{\rho}(\alpha):=[\rho_{S_1}(\alpha),\rho_{S_2}(\alpha),\rho_{C_1}(\alpha),\rho_{C_2}(\alpha)]^\top,
    % \end{equation*}such that the references are given by 
    \begin{equation}\label{eq:rho}
    \begin{bmatrix}
        \dot{r}_{S_1}&\dot{r}_{S_2}&\dot{r}_{C_1}&\dot{r}_{C_2}
    \end{bmatrix}^\top = \boldsymbol{\rho}(\alpha,f_\alpha),
    \end{equation}
    where $\alpha \in [0,2\pi)$ represents the commutation angle governing the relative progression through a step of the actuator. The angle $\alpha$ is determined by a user-defined drive frequency $f_\alpha(t)$ that specifies the number of steps per second:
    \begin{equation}
        \alpha(t) = \alpha(t_0) +  2\pi\operatorname{mod}\left( \int_{t_0}^t f_\alpha(\tau)\textnormal{d}\tau,1\right).
            \end{equation}
            Figure~\ref{fig:control_scheme} illustrates this open-loop control scheme, where the feedforward controller determines the piezo voltages based on $\alpha$. The piezo elements then produce individual displacements, which collectively drive the mover through the kinematics $\kappa$. However, obtaining accurate hysteresis models $\hat{M}_e$ is challenging, as they must $(i)$ be invertible, $(ii)$ allow real-time evaluation at fast sampling rates, and $(iii)$ correctly capture rate-dependent hysteresis effects. Any imperfections in the hysteresis model lead to a position error of the mover, degrading performance. The next section describes another parasitic effect that deteriorates positioning performance, which is not addressed by hysteresis compensation.

\subsection{$\alpha$-dependent disturbances}\label{sec:alpha-dependent}
The second parasitic effect arises from \(\alpha\)-dependent disturbances caused by slip and misalignments. Experimental data of the position error \(\varepsilon(t_k) = r(t_k) - y(t_k)\) for a range of constant drive frequencies is shown in Figure~\ref{fig:alpha-dependent}. These experiments use simple reference displacements \(\boldsymbol{\rho}(\alpha,f_\alpha)\), as displayed in Figure~\ref{fig:waveforms}, and a constant hysteresis model \begin{equation}\label{eq:hystconst}
    \hat{M}_{e,\text{const}}=c_{M,e},\quad \forall e\in\Omega, \quad c_{M,e}\in \mathbb{R}.
    \end{equation}
Figure~\ref{fig:alpha-dependent} demonstrates that the tracking error is consistent in the \(\alpha\)-domain across steps, significantly degrading positioning performance. This suggests that modifying \(\boldsymbol{\rho}(\alpha,f_\alpha)\) could improve positioning accuracy, as \(\boldsymbol{\rho}(\alpha,f_\alpha)\) governs the repeating motion of the piezoelectric elements. Variations over drive frequencies primarily result from lowpass effects of the capacitive sensor, as discussed in detail in Section~\ref{sec:experimental}. Moreover, the data indicates that the $\alpha$-dependent disturbances are direction-dependent, likely because a misalignment would cause a positive force in one direction, along the motion of the mover, but a negative force in the other direction, counteracting the motion. 

\subsection{Problem definition} 
The aim is to design a feedforward control law \begin{equation}
    u_e(t_k)=u_e(t_0) + f(r(t_k),\ldots,r(t_{0})), \quad \forall e\in\Omega
\end{equation} that minimizes the root-mean-square deviation (RMSD) of the mover position error \(\varepsilon(t_k) = r(t_k) - y(t_k)\) for arbitrary reference signals \(r(t_k)\), where the RMSD is given by
\begin{equation}\label{eq:rmsd}
    \varepsilon_{\text{RMSD}}=\sqrt{\frac{1}{N} \sum_{k=1}^N \left(\varepsilon(t_k) - \frac{1}{N}\sum_{k=1}^N \varepsilon(t_k)\right)^2}.
\end{equation} 
This measure quantifies the velocity ripples of the piezo-stepper actuator when the reference velocity is constant, but the formulation remains applicable to varying reference velocities as well. Two sub-problems are defined:
\begin{enumerate}
    \item {Hysteresis compensation:} develop and invert hysteresis models \(\hat{M}_e\) to ensure accurate tracking \(\dot{y}_e \approx \dot{r}_e\) despite the absence of direct piezo displacement measurements.
    \item Compensation of $\alpha$-domain disturbances: design \(\boldsymbol{\rho}(\alpha,f_\alpha)\) to minimize $\varepsilon_{\text{RMSD}}$, independent of drive frequency \(f_\alpha\), by leveraging mover position data \(y\). 
    \end{enumerate}
    The next section provides the solution to the first sub-problem, and Section~\ref{sec:misalignments} addresses the second sub-problem.

\section{Rate-dependent hysteresis compensation}\label{sec:hysteresis}
This section details the developed approach to rate-dependent hysteresis compensation in piezo elements. 
\subsection{Overview}
First, an overview of the method is given, starting with a description of the control law governing the scheme in Figure~\ref{fig:control_scheme}. To realize a desired movement $r_e$ of a single piezo element $e$, integration of~\eqref{eq:hysteresis} leads to
 \begin{equation}\label{eq:control_ct}
    u_e(t) = u_e(t_0) + \int_{t_0}^t \frac{\dot{r}_e(\tau)}{M_e(\dot{u}_e(\tau),u_{e,{a}}(\tau))} d \tau,
\end{equation} 
where \(u_e\) is the voltage required to achieve \(y_e = r_e\), and \(u_{e,a}(t)\) is the absement~\cite{Strijbosch2023,JSpei2018}, which quantifies the accumulated voltage change since the last reversal:
\begin{equation}
    u_{e,a}(t) := |u_e(t) - u_e(t_{\text{turn}})|, \quad t \geq t_{\text{turn}},
\end{equation}
with \(t_{\text{turn}} = \max_{\tau}\{\tau < t : \dot{u}_e(\tau)\dot{u}_e(\tau + \epsilon) < 0\}\), where \(\epsilon\) is an infinitesimally small positive number. Applying first-order discretization of~\eqref{eq:control_ct} and introducing a delay in \(\dot{u}_e(\tau)\) to avoid algebraic loops leads to the following control law:
\begin{equation}\label{eq:controllaw}
u_e(t_k) = u_e(t_{k-1}) + \frac{T_s \dot{r}_{e}(t_{k-1})  }{\hat{M}_{\boldsymbol{\theta}_{e}}\left(\frac{|u_e(t_{k-1})-u_e(t_{k-2})|}{T_s},u_{e,a}(t_{k-1})\right)}, \quad k\geq 1, 
\end{equation}
where \(\boldsymbol{\theta}_e\) are the hysteresis model parameters, $u(t_k)=0$ for $k\leq 0$, and $\dot{r}_e(t_k)$ is defined by~\eqref{eq:rho}. %Moreover, $g_e(f_\alpha)$ is a scaling factor detailed later. %Moreover, superscript $\iota$ accounts for direction-dependent hysteresis as detailed later, and is defined by \begin{equation}
% \iota:= \begin{cases}
%     + & \text{if } \dot{r}_e(t_k) \geq 0,\\
%     - & \text{if } \dot{r}_e(t_k) < 0.
% \end{cases}
% \end{equation} 
This control law comprises two contributions explained in the remainder of this section:
\begin{enumerate}
    \item {Identification of a hysteresis model:} a model \(\hat{M}_{\boldsymbol{\theta}_e}\) is identified from measurements of the currents to approximate \(M_e\), as explained in Section~\ref{sec:model}.
    \item {Implementation aspects:} practical aspects of implementing the control law, such as a computationally efficient implementation and a solution to integrator drift, are addressed in Section~\ref{sec:impl_hyst}.
\end{enumerate}

\subsection{Modeling rate-dependent hysteresis}\label{sec:model}
The hysteresis function $M_e$ in~\eqref{eq:hysteresis} describes the relationship between piezo voltages and their displacements. While the voltages are known, being the control input, the individual piezo displacements are not measured directly. Instead, they are estimated from measurements of the current as follows. The rate of displacement of a piezoelectric element, neglecting changes in external forces over time, is given by~\cite{Fang2013} 
    \begin{equation}\label{eq:current}
        \dot{y}_e(t) = \xi_e i_e(t),
        \end{equation}
where $i_e$ is the measured current through element $e$ and $\xi_e$ a constant in ms$^{-1}$A$^{-1}$. The key idea is to perform experiments with varying voltage signals to obtain measurements of \begin{equation}\label{eq:M_observe}
    \frac{1}{\xi_e} M_e(\dot{u}_e, u_{e,{a}}) =\left|\frac{i_e}{\dot{u}_e}\right|.
\end{equation}
Since the model is inverted in the control law, the value of $\xi_e$ has no practical significance. These measurements result in datasets
\begin{equation}
    \mathcal{D}_e = \{\dot{u}_e(t_k), u_{e,a}(t_k), i_e(t_k)\}_{k=1}^n, 
\end{equation}
for all $e\in\Omega$. To account for direction-dependent hysteresis effects, separate models $\hat{M}_{\theta_{e^+}}$ and $\hat{M}_{\theta_{e^-}}$ are fit on this dataset, which are then used in the feedforward control law, such that \begin{equation}\label{eq:M_dirdep}
\hat{M}_{\theta_e}=\begin{cases}
    \hat{M}_{\theta_{e^+}} & \text{if } \dot{r}_e(t_k) \geq 0,\\
    \hat{M}_{\theta_{e^-}} & \text{if } \dot{r}_e(t_k) < 0.
\end{cases}
\end{equation}
The next section explains the design of a single experiment that yields all datasets $\mathcal{D}_{e}$. Subsequently, the chosen model structure and fitting method are detailed. 

% resulting in a data-set $\mathcal{D}_e=\{\dot{u}_e(t_k), u_{e,a}(t_k),i_e(t_k)\}_{k=1}^n$ for all $e\in\Omega$, after which a model $\hat{M}_e\approx M_e$ is fit. This model is inverted in the feedforward control law, so the value of $\xi_e$ has no practical significance. The next section explains the design of experiments that yield $\mathcal{D}_e$. Subsequently, the chosen model structure and fitting method are detailed. 

\subsubsection{Experiment design}
For the datasets $\mathcal{D}_{e}$ to be useful in modeling  $\hat{M}_{\theta_{e^+}}$ and $ \hat{M}_{\theta_{e^-}}$, the measured voltage pairs \((\dot{u}_e(t_k), u_{e,a}(t_k)\)) should cover a range relevant to the application. Under control law~\eqref{eq:controllaw} with some imperfect $\hat{M}_e$ and ${\rho}_e$ that achieve a stepping motion, albeit with parasitic effects, all pairs \((\dot{u}_e(t_k), u_{e,a}(t_k)\)) depend solely on the chosen drive frequency $f_\alpha$. Therefore, a grid $\mathcal{F}$ of $F$ positive and negative drive frequencies in an application-relevant range between $f_{\alpha,\min}$ and $f_{\alpha,\max}$ is defined as \begin{equation}\label{eq:Fgrid}
    \mathcal{F} = \left\{ \pm f_{\alpha,i} \mid f_{\alpha,i} = f_{\alpha,\min} \left(\frac{f_{\alpha,\max}}{f_{\alpha,\min}}\right)^{\frac{i-1}{F-1}}, \quad i = 1, \dots, F \right\},
    \end{equation}
    where logarithmic spacing is used because of the potentially wide range of drive frequencies. Section~\ref{sec:experimental} addresses the choice of $F$. An experiment is started where the drive frequency cycles stepwise through $\mathcal{F}$ for $n_s$ steps per frequency, with control law~\eqref{eq:controllaw}, using a constant hysteresis model~\eqref{eq:hystconst} and the following nominal waveform function $\boldsymbol{\rho}(\alpha,f_\alpha)$ used in literature~\cite{Aarnoudse2023}, defined as
\begin{equation}\label{eq:waveforms}
    \begin{aligned}
        {\rho}_{C_i}(\alpha,f_\alpha) =& \begin{cases}
             |f_\alpha|\frac{r_{C_i,\max}-r_{C_i,\min}}{2\pi/3}(2i-3)    & 0 \leq \alpha < \frac{\pi}{3},\\
            0                                                        & \frac{\pi}{3} \leq \alpha < \frac{2\pi}{3},\\
            - |f_\alpha|\frac{r_{C_i,\max}-r_{C_i,\min}}{2\pi/3}(2i-3)         & \frac{2\pi}{3} \leq \alpha < \frac{4\pi}{3},\\
            0                                                        & \frac{4\pi}{3} \leq \alpha < \frac{5\pi}{3},\\
            |f_\alpha|\frac{r_{C_i,\max}-r_{C_i,\min}}{2\pi/3}(2i-3)   & \frac{5\pi}{3} \leq \alpha < 2\pi,
        \end{cases}\\
        {\rho}_{S_1}(\alpha,f_\alpha) =& \begin{cases}
            |f_\alpha|\frac{r_{S_1,\max}-r_{S_1,\min}}{5\pi/3}    & 0 \leq \alpha < \frac{\pi}{3},\\
            -|f_\alpha|\frac{r_{S_1,\max}-r_{S_1,\min}}{\pi/3}    & \frac{\pi}{3} \leq \alpha < \frac{2\pi}{3},\\
            |f_\alpha|\frac{r_{S_1,\max}-r_{S_1,\min}}{5\pi/3}         & \frac{2\pi}{3} \leq \alpha < 2\pi,
        \end{cases}\\
        {\rho}_{S_2}(\alpha,f_\alpha) =& \begin{cases}
            |f_\alpha|\frac{r_{S_2,\max}-r_{S_2,\min}}{5\pi/3}    & 0 \leq \alpha < \frac{4\pi}{3},\\
            - |f_\alpha|\frac{r_{S_2,\max}-r_{S_2,\min}}{\pi/3}    & \frac{4\pi}{3} \leq \alpha < \frac{5\pi}{3},\\
            |f_\alpha|\frac{r_{S_2,\max}-r_{S_2,\min}}{5\pi/3}         & \frac{5\pi}{3} \leq \alpha < 2\pi,
        \end{cases}
    \end{aligned}
\end{equation}
where $i\in\{1,2\}$ and $r_{e,\max},\ r_{e,\min}$ define the maximum reference stroke of the elements. These reference displacements, depicted in Figure~\ref{fig:waveforms}, lead to similarly shaped voltage waveforms when scaled with a scalar \(\hat{M}_{e,\text{const}}\) through control law~\eqref{eq:controllaw}. Hence, for every frequency $f_{\alpha,i}\in\mathcal{F}$, the measurements include data of one negative voltage rate, one positive voltage rate, and a range of different voltage absements. %Note that the choice $g_e(f_\alpha)=1$ leads to a decrease in stroke for higher drive frequencies. This is intentional as it yields a large range of voltage rates, which facilitates the modeling of $\hat{M}$; Section~\ref{sec:stroke} introduces a scaling factor $g_e(f_\alpha)$ that renders the stroke independent of drive frequency. 
Algorithm~\ref{algo:datacol} summarizes the data collection process, and the next section describes the hysteresis model structure. 

\begin{algorithm}[tb]
\caption{Data collection for hysteresis compensation}\label{algo:datacol}
\begin{algorithmic}[1]
\Require Range \(f_{\alpha,\min},\ldots,f_{\alpha,\max}\) of $F$ application-relevant drive frequencies, number of steps $n_s$. 
\State Initialize  $\mathcal{D}_e=\{ \}$ for all $e\in\Omega$, define $\mathcal{F}$ as in~\eqref{eq:Fgrid}.
\State Start experiment with control law~\eqref{eq:controllaw} using constant hysteresis model~\eqref{eq:hystconst}, $f_\alpha=0$ and waveform function~\eqref{eq:waveforms}.
\For{$f_{\alpha,i}\in\mathcal{F}$}
        \State Update $f_\alpha\leftarrow f_{\alpha,i}$.
        \State Wait for $t_{\text{step}}=\frac{n_s}{f_\alpha}$ seconds.
\EndFor
\State Stop the experiment and store all $\left(\dot{u}_e(t_k), u_{e,a}(t_k),i_e(t_k)\right)$ in $\mathcal{D}_e$ for all $e\in\Omega$.
\State \Return Data-sets $\mathcal{D}_e$, $\forall e\in\Omega$.
    \end{algorithmic}
\end{algorithm}

\subsubsection{Model structure}
To capture the rate-dependent hysteresis effects, the model structure must account for both the voltage rate \(\dot{u}_e\) and the absement \(u_{e,a}\). The Ramberg-Osgood model~\cite{Ramberg1943}, which has proven successful in modeling hysteresis for fixed drive frequencies, does not include rate dependency~\cite{Strijbosch2023}. Therefore, an alternative structure is presented that, like the Ramberg-Osgood model, is smooth in the voltage absement, but also allows for rate-dependency. 
The model is parametrized linearly in the parameters as 
\begin{equation}\label{eq:hyst_model}
    \hat{M}_{\boldsymbol{\theta}_{e^{\iota}}}(\mathbf{x}_e) = \mathbf{k}^\top(\mathbf{x}_e) \boldsymbol{\theta}_{e^{\iota}},\quad \iota\in\{+,-\},
    \end{equation}
    where the input vector is given by
    \begin{equation}
        \mathbf{x}_e := [|\dot{u}_e|, u_{e,a}]^\top.
    \end{equation}
    Here, \(\mathbf{k}(\mathbf{x}_e)\) is a vector of basis functions evaluated at the input \(\mathbf{x}_e\), given by
    \begin{equation}
        \mathbf{k}(\mathbf{x}_e) = [k(\mathbf{x}_e, \mathbf{x}_{e,1}), \dots, k(\mathbf{x}_e, \mathbf{x}_{e,m})]^\top,
    \end{equation}
    where \(\mathbf{x}_{e,i}, i \in \{1,\ldots,m\}\), are predefined grid points in \(\mathbb{R}^2\), and \(k:\mathbb{R}^2 \times \mathbb{R}^2\to \mathbb{R}\) is a kernel function defining the structure of the model. To account for direction-dependent hysteresis effects, separate parameter vectors \(\boldsymbol{\theta}_{e^+}\) and \(\boldsymbol{\theta}_{e^-}\) are used for positive and negative velocity directions, see~\eqref{eq:M_dirdep}. %Since the input \(\mathbf{x}_e\) is defined using \(|\dot{u}_e|\), both models share the same input representation while being independently parameterized, allowing them to capture direction-dependent hysteresis behavior.
    % for any arbitrary $\mathbf{x}_e:=[|\dot{u}_e|,u_{e,{a}}]^\top$. Here, $\mathbf{k}(\mathbf{x}_e):\mathbb{R}^2\to \mathbb{R}^{m\times 1}$ is given by \begin{equation}
    % \mathbf{k}(\mathbf{x}_e) = [k(\mathbf{x}_e,\mathbf{x}_{e,1}),\ldots,k(\mathbf{x}_e,\mathbf{x}_{e,m})]^\top,
    % \end{equation}
    % where $\mathbf{x}_{e,i},\ i\in\{1,\ldots,m\}$ form a user-defined grid of points in $\mathbb{R}^2$ and $k:\mathbb{R}^2 \times \mathbb{R}^2\to \mathbb{R}$ a kernel function that defines the model structure. 

    An example basis function that is useful in modeling functions when the nonlinear structure is smooth but not fully known, popularized by Gaussian Process regression~\cite{Rasmussen2004}, is given by  \begin{equation}\label{eq:kernel}
    k(\mathbf{x}_e,\mathbf{x}_e') = \sigma_f^2\exp\left(-\frac{1}{2}\left(\mathbf{x}_e-\mathbf{x}_e'\right)^\top \boldsymbol{\Sigma}^{-1}\left(\mathbf{x}_e-\mathbf{x}_e'\right)\right).
    \end{equation}
    Here, $\boldsymbol{\Sigma}=\operatorname{diag}(\ell_1,\ell_2)$, $\ell_1,\ell_2\in \mathbb{R}$ are tuning parameters that govern the smoothness of the basis functions in the direction of $\dot{u}_e$ and $u_{e,{a}}$, respectively. The parameter $\sigma_f^2$ scales the overall magnitude of the basis functions. To avoid the need to evaluate these kernel functions online, Section~\ref{sec:impl_hyst} presents a resource-efficient implementation. 
 
\subsubsection{Obtaining the model}
The parameters $\boldsymbol{\theta}_{e^{\iota}}$ in~\eqref{eq:hyst_model} are obtained by solving the following linear least-squares problem for each velocity direction \(\iota \in \{+, -\}\):\begin{equation}
    \min_{\boldsymbol{\theta}_{e^{\iota}}} \| \mathbf{K}(\mathbf{X}_e^{\iota}) \boldsymbol{\theta}_{e^{\iota}} - \mathbf{m}_e^{\iota} \|_2^2,
\end{equation}
where the data matrices are constructed separately for positive and negative velocity directions:
\begin{subequations} 
    \begin{align}
    \mathbf{X}_e^{+} &= \{ \mathbf{x}_e(t_k) \mid \dot{u}_e(t_k) \geq 0, \mathbf{x}_e(t_k),\dot{u}_e(t_k) \in \mathcal{D}_e \}, \\
    \mathbf{X}_e^{-} &= \{ \mathbf{x}_e(t_k) \mid \dot{u}_e(t_k) < 0, \mathbf{x}_e(t_k),\dot{u}_e(t_k) \in \mathcal{D}_e \}.
    \end{align}
\end{subequations}
The corresponding kernel matrices and measurement vectors are:
\begin{align}
    \mathbf{K}({\mathbf{X}_e^{\iota}}) &= \big[ \mathbf{k}({\mathbf{x}}_{e,1}^{\iota}),\ldots,\mathbf{k}({\mathbf{x}}_{e,n^{\iota}}^{\iota}) \big]^\top \in\mathbb{R}^{n^{\iota} \times m}, \\
    \mathbf{m}_e^{\iota} &= \begin{bmatrix}
    \left|\frac{i_e(t_1)}{\dot{u}_e(t_1)}\right| & \ldots & \left|\frac{i_e(t_{n^{\iota}})}{\dot{u}_e(t_{n^{\iota}})}\right|
    \end{bmatrix}^\top.
\end{align}
The least-squares solution for each direction is then given by
\begin{equation}\label{eq:hysteresis_fit}
    \hat{\boldsymbol{\theta}}_{e^{\iota}} = \big(\mathbf{K}^\top({\mathbf{X}_e^{\iota}}) \mathbf{K}({\mathbf{X}_e^{\iota}}) \big)^{-1} \mathbf{K}^\top({\mathbf{X}_e^{\iota}}) \mathbf{m}_e^{\iota}.
\end{equation}
This solution is valid as long as \(\mathbf{K}^\top({\mathbf{X}_e^{\iota}}) \mathbf{K}({\mathbf{X}_e^{\iota}})\) is full rank, which is typically satisfied since \(m \ll n^{\iota}\). 

Thus, the obtained model \(\hat{M}_{\hat{\boldsymbol{\theta}}_e}\) captures rate-dependent and direction-dependent hysteresis effects and enables online compensation. The next section addresses implementation aspects.

% with $\mathbf{K}({\mathbf{X}_e})=\allowbreak[\mathbf{k}({\mathbf{x}}_{e,1}),\ldots,\mathbf{k}({\mathbf{x}}_{e,n})]^\top$ $\in\mathbb{R}^{n\times m}$, ${\mathbf{X}_e}=$ $[{\mathbf{x}}_{e,1}^\top,\ldots,{\mathbf{x}_e}_n^\top]^\top \allowbreak\in \mathcal{D}$, and \begin{equation}
% \mathbf{m}_e = \begin{bmatrix}
% \left|\frac{i_e(t_1)}{\dot{u}_e(t_1)}\right| & \ldots & \left|\frac{i_e(t_n)}{\dot{u}_e(t_n)}\right|
% \end{bmatrix}^\top \in\mathcal{D}_e.
% \end{equation} The solution is given by \begin{equation}\label{eq:hysteresis_fit}
%     \hat{\boldsymbol{\theta}}_{e^{\iota}} = \left(\mathbf{K}^\top({\mathbf{X}_e}) \mathbf{K}({\mathbf{X}_e})\right)^{-1} \mathbf{K}^\top({\mathbf{X}_e}) \mathbf{m}_e.
% \end{equation}
% This solution is valid only if $\mathbf{K}^\top({\mathbf{X}_e}) \mathbf{K}({\mathbf{X}_e})$ is full rank, which is not a problem in practice, since $m\ll n$. Hence,~\eqref{eq:hyst_model} and~\eqref{eq:hysteresis_fit} provide a hysteresis model $\hat{M}_{\hat{\boldsymbol{\theta}}_e}$ that allows for online hysteresis compensation. The next section addresses some implementation aspects.

\subsection{Implementation aspects}\label{sec:impl_hyst}
This section covers some implementation aspects, starting with a resource-efficient implementation of the hysteresis model. 
\subsubsection{Resource-efficient implementation}
The control law~\eqref{eq:controllaw} requires the evaluation of the hysteresis model $\hat{M}_{\boldsymbol{\theta}_{e^{\iota}}}$ at high sampling rates, which is computationally challenging because of the nonlinear terms in~\eqref{eq:kernel}. Therefore, $\hat{M}_{\boldsymbol{\theta}_{e^{\iota}}}$ is used to fill a lookup table $\hat{M}_{e^{\iota}}^{\text{LUT}}(\mathbf{x}_e)$ offline. This renders the approach feasible for embedded hardware with large memory but limited computational capabilities, only requiring simple computations for linear interpolation:
 \begin{equation}\begin{aligned}
\hat{M}_{e^{\iota}}^{\text{LUT}}(\mathbf{x}_e) =& \begin{bmatrix}
    1-\tau_1 & \tau_1
\end{bmatrix} \mathbf{Y}^{\iota} \begin{bmatrix}
    1-\tau_2\\ \tau_2
\end{bmatrix},\\
\tau_1 =& \frac{x_1-x_1^{(i)}}{x_1^{(i+1)}-x_1^{(i)}},\quad \tau_2=\frac{x_2-x_2^{(j)}}{x_2^{(j+1)}-x_2^{(j)}},\\
\mathbf{Y}^{\iota} =&\begin{bmatrix}
    \hat{M}_{\boldsymbol{\theta}_{e^{\iota}}}\left(\begin{bmatrix}x_1^{(i)}\\ x_2^{(j)}\end{bmatrix}\right) & \hat{M}_{\boldsymbol{\theta}_{e^{\iota}}}\left(\begin{bmatrix}x_1^{(i)}\\ x_2^{(j+1)}\end{bmatrix}\right) \\
    \hat{M}_{\boldsymbol{\theta}_{e^{\iota}}}\left(\begin{bmatrix}x_1^{(i+1)}\\ x_2^{(j)}\end{bmatrix}\right) & \hat{M}_{\boldsymbol{\theta}_{e^{\iota}}}\left(\begin{bmatrix}x_1^{(i+1)}\\ x_2^{(j+1)}\end{bmatrix}\right)
\end{bmatrix}.
\end{aligned}
\end{equation}
Here, $(x_1^{(i)},x_2^{(j)},x_1^{(i+1)},x_2^{(j+1)})$ are grid points surrounding $\mathbf{x}_e$, satisfying \begin{equation}
x_1^{(i)}\leq x_1 \leq x_1^{(i+1)},\quad x_2^{(j)}\leq x_2\leq x_2^{(j+1)},
\end{equation} 
and $\mathbf{Y}^{\iota}$ is computed offline for both $\iota\in\{+,-\}$. The next section addresses the issue of integrator drift in the control law.

\subsubsection{Integrator drift}\label{sec:drift}
While reference displacement rates~\eqref{eq:waveforms} are zero-mean, the division of these references by rate-dependent hysteresis model $\hat{M}_e$ in~\eqref{eq:controllaw} may lead to a voltage rate that is not zero-mean, leading to drift in the voltage $u_e$. In the worst case, this could lead to the clamps not fully extending to press the shears onto the mover, impairing stepping performance. 

To mitigate the accumulation of this integrator drift over time, an anti-windup mechanism is introduced by constraining the output whenever the piezo voltages exceed user-defined bounds $u_{e,\min}$ and $u_{e,\max}$. These artificial bounds are defined for the purpose of drift compensation only and are stricter than the amplifier limits, i.e., $u_{e,\min}>u_{e,\min,\text{amp}}$ and $u_{e,\max}<u_{e,\max,\text{amp}}$. Hence, the final control law is given by \begin{equation}\label{eq:controllaw2}
\begin{aligned}
    u_e(t_k) =& \begin{cases}
    u_{e,\min} & \text{if } \upsilon< u_{e,\min},\\
    u_{e,\max} & \text{if } \upsilon > u_{e,\max},\\
    \upsilon & \text{otherwise}.
\end{cases}\\
\upsilon  =&  u_e(t_{k-1}) + \frac{T_s \dot{r}_e(\alpha(t_{k-1}))}{\hat{M}_{{e}}^{\text{LUT}}\left(\left[\frac{|u_e(t_{k-1})-u_e(t_{k-2})|}{T_s},u_{e,a}(t_{k-1})\right]^\top\right)}.
\end{aligned}
\end{equation}
As long as the piezo voltages remain within the bounds $u_{e,\min}$ and $u_{e,\max}$, the drift compensation is inactive, and integrator drift can still occur, deteriorating stepping performance. On the other hand, when the bounds are reached, the control law saturates the output, preventing further drift but also limiting the achievable stroke. Therefore, it is desired that the voltage waveforms exceed the bounds at least once per cycle of the drive frequency. This condition is formally expressed as:
\begin{equation}\label{eq:saturate}
\begin{aligned}
    u_e(t_{\text{top}}) &> u_{e,\max}, \quad \forall t_{\text{top}} \in \mathcal{T}_{\text{top}}(f_\alpha),\\
    u_e(t_{\text{bot}}) &< u_{e,\min}, \quad \forall t_{\text{bot}} \in \mathcal{T}_{\text{bot}}(f_\alpha),
\end{aligned}
\end{equation}
where the sets \( \mathcal{T}_{\text{top}}(f_\alpha) \) and \( \mathcal{T}_{\text{bot}}(f_\alpha) \) identify the time instances at which \( u_e \) reaches its maximum and minimum values within one period \( T = 1/f_\alpha \):
\begin{equation}\label{eq:topbot}
    \begin{aligned}
        \mathcal{T}_{\text{top}}(f_\alpha) &=\left\{ t \in [0, T] : u_e(t) = \max_{t \in [0, T]} u_e(t) \right\}, \\
        \mathcal{T}_{\text{bot}}(f_\alpha) &= \left\{ t \in [0, T] : u_e(t) = \min_{t \in [0, T]} u_e(t) \right\}.
    \end{aligned}
\end{equation} 
This condition is satisfied by appropriate selection of reference strokes \( r_{e,\max} \) and \( r_{e,\min} \) in~\eqref{eq:waveforms}. These bounds are determined \textit{offline} through optimization, once after Algorithm~\ref{algo:datacol}, ensuring that for each selected \( f_\alpha \), the voltage \( u_e(t) \) exceeds the artificial limits at least once per cycle. 
The optimization problem is given by \begin{equation}\begin{aligned}\label{eq:opt}
        \min_{{r}_{e,\max},{r}_{e,\min}} & J_{e,f_\alpha}(\dot{r}_{e,\max},\dot{r}_{e,\min}) \\
        = & \sum_{t_{\text{top}}\in \mathcal{T}_{\text{top},2}(f_\alpha)}\|u_e(t_{\text{top}})-u_{e,\max}\|_2^2 \\
        & + \sum_{t_{\text{bot}}\in \mathcal{T}_{\text{bot},2}(f_\alpha)}\|u_e(t_{\text{bot}})-u_{e,\min}\|_2^2,\\
        \text{subject to } & \text{hysteresis model}~\eqref{eq:hyst_model},\text{ waveforms}~\eqref{eq:waveforms},\\
        & \text{ control law}~\eqref{eq:controllaw2},\\
        & \alpha(t_k) = 2\pi \text{mod}\left(\alpha(t_{k-1}) + T_s f_\alpha\right), \quad \forall k \in \mathcal{K}_{\text{sim},}\\
        & u_e(t_0) = \alpha(t_0) = 0,\\
        & \mathcal{T}_{(\cdot),2}(f_\alpha) = \mathcal{T}_{(\cdot)}(f_\alpha/2) \cap \{t \mid t\leq {1}/{f_\alpha}\}.
    \end{aligned}
\end{equation}
Here, the simulated indices are \( \mathcal{K}_{\text{sim}}=\{1,\ldots,2\lfloor F_s / f_\alpha \rfloor\} \), with flooring operator $\lfloor\cdot\rfloor$, consisting of two periods to account for transients. The problem is solved using interior-point optimization with approximate gradients~\cite{nocedal1999}. 

The resulting optimal bounds \( r_{e,\max}^\star \) and \( r_{e,\min}^\star \) are precomputed for different drive frequencies and stored in lookup tables:
\begin{equation}\label{eq:g}
    \begin{aligned}
        {r}_{e,\max}^\star &= g_{e,\max}^{\text{LUT}}(f_\alpha),\\
        {r}_{e,\min}^\star &= g_{e,\min}^{\text{LUT}}(f_\alpha).
    \end{aligned}
\end{equation}
The reference movements~\eqref{eq:waveforms}, combined with control law~\eqref{eq:controllaw2} and these optimized bounds, ensure continuous and repeatable motion of the mover despite hysteresis, while preventing drift due to modeling errors. However, they do not yet compensate for slip or mechanical misalignments, which are addressed in the next section.

\section{Compensation of $\alpha$-dependent disturbances}\label{sec:misalignments}
The control law~\eqref{eq:controllaw2} enables accurate tracking of reference movements $\dot{r}_e(t)$ of individual piezo elements by compensating hysteresis, which is the first parasitic effect. The second parasitic effect, consisting of mechanical misalignments and slip, introduces velocity ripples that are repetitive in the $\alpha$-domain. 
This section introduces a data-driven approach to compensate for these $\alpha$-dependent disturbances, starting with an overview of the method.

\subsection{Overview}
The developed approach compensates for $\alpha$-domain disturbances by iteratively learning a compensation function that modifies the shear waveforms. These disturbances, shown in Figure~\ref{fig:alpha-dependent}, are repetitive in the $\alpha$-domain and can therefore be mitigated effectively using Iterative Learning Control (ILC)~\cite{Aarnoudse2023}. Only the shear waveforms are modified as these directly affect the mover displacement through~\eqref{eq:mover}. The approach is summarized in Algorithm~\ref{algo:alpha} and involves three contributions:
\begin{enumerate}
    \item A compensation function $\dot{f}_{\boldsymbol{\gamma}}^{\text{proj}}(\alpha)$ is parametrized to modify the shear waveforms. This function is updated iteratively using ILC to minimize the effects of $\alpha$-domain disturbances, as detailed in Section~\ref{sec:compensation}.
    \item Monotonic convergence of the compensation function over trials is analyzed in Section~\ref{sec:convergence}.
    \item Implementation aspects such as design of the learning filter 
    and direction-dependency of the compensation function are described in Section~\ref{sec:impl_alpha}.
\end{enumerate}
In contrast to~\cite{Aarnoudse2023}, the approach presented in this paper integrates compensation of $\alpha$-dependent disturbances with rate-dependent hysteresis compensation of all piezo elements, such that the learned compensation function is applicable to arbitrary drive frequencies and does not require relearning when the task changes.

\begin{algorithm}[tb]
    \caption{Compensation of $\alpha$-domain disturbances}\label{algo:alpha}
    \begin{algorithmic}[1]
    \Require Hysteresis models $\hat{M}_{{e}}^\text{LUT}$, ${e}\in{\Omega}$, functions $g_{e,\max}^{\text{LUT}}(f_\alpha),g_{e,\min}^{\text{LUT}}(f_\alpha)$, LTI model $\hat{G}(q)$, drive frequency $f_\alpha$.
    \State Design filters $L(q)$, $Q(q)$, see Section~\ref{sec:lti_design}.
    \State Initialize $\boldsymbol{\gamma}_0=\mathbf{0}$.
    \For{$j \in [0,\ldots,n_{\text{tr}}-1]$}
    \State Conduct an experiment with control law~\eqref{eq:controllaw2}, using reference~\eqref{eq:ff}. Store $y_j$.
    \State Compute $\dot{f}_{j+1}(t_k)$ with~\eqref{eq:update}.
    \State Compute $\boldsymbol{\gamma}_{j+1}$ with~\eqref{eq:solproj}. 
    \EndFor
    \State \Return the final compensation function $\dot{{f}}_{\boldsymbol{\gamma}_{n_{\text{tr}-1}}}^{\text{proj}}(\alpha)$.
        \end{algorithmic}
    \end{algorithm}

\subsection{Learning a compensation function}\label{sec:compensation}
The compensation function $\dot{f}_{\boldsymbol{\gamma}}^{\text{proj}}(\alpha)$ modifies the shear reference displacements to counteract the parasitic effects in the $\alpha$-domain, as given by:
\begin{equation}\label{eq:ff}
    \dot{\tilde{r}}_{S_i}(\alpha) = \dot{r}_{S_i}(\alpha) + \dot{f}_{\boldsymbol{\gamma}}^{\text{proj}}(\alpha),
    \end{equation}
    where $\dot{r}_{S_i}(\alpha)$ is the nominal reference defined in~\eqref{eq:waveforms}, and $\dot{f}_{\boldsymbol{\gamma}}^{\text{proj}}(\alpha)$ is a compensation function with parameters $\boldsymbol{\gamma}$. 
    When these modified references $\dot{\tilde{r}}_{S_i}(\alpha)$ are used in control law~\eqref{eq:controllaw2} with $\hat{M}_e\approx M_e$, this results in $y_{e}\approx \tilde{r}_{e}$, i.e., the motion of the shears approximately equals the modified references. 
    The goal of the compensation function is then to minimize the tracking error of the mover: 
    \begin{equation}\label{eq:error}
        \varepsilon_j(t_k) := r(t_k)-y_j(t_k),
        \end{equation}
         where $j$ is a trial number, and the reference is defined as \begin{equation}
        r(t_k) := \hat{G}(q) \dot{r}_{S_i}(t_k),
        \end{equation}
        with $\hat{G}(q)\approx G(q)$ a model detailed in Section~\ref{sec:G}. The next sections detail how the compensation function $\dot{f}_{\boldsymbol{\gamma}}^{\text{proj}}(\alpha)$ is constructed from data over several trials to minimize $\|\varepsilon_j\|_2$, starting with the parametrization of $\dot{f}_{\boldsymbol{\gamma}}^{\text{proj}}(\alpha)$.

    \subsubsection{Parametrization of the compensation function} 
    The compensation function $\dot{f}_{\boldsymbol{\gamma}}^{\text{proj}}(\alpha)$ is parametrized as a piecewise-linear function to accommodate for the limited computational resources available on control hardware. First, define a grid of $n_\gamma$ points $\alpha_c\in [0,2\pi)$, each equidistantly spaced $\Delta_\gamma$ apart. Next, define \begin{equation}\label{eq:basis}
        \dot{f}_{\boldsymbol{\gamma}}^{\text{proj}}(\alpha) = \boldsymbol{\psi}^\top (\alpha) \boldsymbol{\gamma},
    \end{equation}
    where $\boldsymbol{\psi}(\alpha)\in \mathbb{R}^{n_\gamma}$ is a vector with only two nonzero elements, namely elements $c$ and $c+1$, the two elements on the grid that surround $\alpha$. These two elements of $\boldsymbol{\psi}(\alpha)$ are given by the linear interpolation \begin{equation}
        \begin{aligned}
        \psi_c(\alpha) = \frac{\alpha_{c+1}-\alpha}{\Delta_\gamma},\quad \psi_{c+1}(\alpha) = \frac{\alpha-\alpha_c}{\Delta_\gamma}.
        \end{aligned}
        \end{equation}
        The next section describes how the model parameters are iteratively updated from data.  
    
    \subsubsection{Iteratively learning $\boldsymbol{\gamma}$ from data} 
    The disturbance \(\dot{d}(\alpha)\) is unknown, so the parameters \(\boldsymbol{\gamma}\) cannot be determined directly. Instead, the repetitive nature of \(\alpha\)-domain disturbances is leveraged, updating \(\boldsymbol{\gamma}\) iteratively using measurements of the mover position to reduce the tracking error.

    First, at trial $j=0$ with $\boldsymbol{\gamma}_0=\mathbf{0}$ and zero initial conditions, an experiment is carried out with control law~\eqref{eq:controllaw2} and reference~\eqref{eq:ff}. Next, the system is reset to its initial state and the following update law is applied: \begin{equation}\label{eq:update}
    \dot{f}_{j+1}(t_k) = Q(q) (\dot{f}_{\boldsymbol{\gamma}_j}^{\text{proj}}(t_k)+L(q) \varepsilon_j(t_k)),
    \end{equation}
    where $Q$ and $L$ are LTI filters detailed in Section~\ref{sec:lti_design}. To facilitate the projection of this feedforward signal to the basis $\boldsymbol{\psi}(\alpha)$, a lifted notation is introduced. 
    The model \(\hat{G}(q)\) is represented in finite-time lifted notation by its impulse response matrix \(\hat{\mathbf{G}}\). This matrix captures the input-output relationship over $N$ samples, assuming zero initial conditions, and is constructed from the impulse response coefficients $\hat{g}_k$ as:
\begin{equation}\label{eq:impulse}
\hat{\mathbf{G}} = 
\begin{bmatrix}
\hat{g}_0 & 0 & \cdots & 0 \\
\hat{g}_1 & \hat{g}_0 & \cdots & 0 \\
\hat{g}_2 & \hat{g}_1 & \cdots & 0 \\
\vdots & \vdots & \ddots & \vdots \\
\hat{g}_{N-1} & \hat{g}_{N-2} & \cdots & \hat{g}_0
\end{bmatrix}.
\end{equation}
This structure generalizes to any linear time-invariant (LTI) system but is here used specifically for \(\hat{G}(q)\). Using this notation, the time-domain signal \(\dot{f}_{\boldsymbol{\gamma}_{j+1}}(t_k)\) obtained from~\eqref{eq:update} is projected to the low-order basis~\eqref{eq:basis} by minimizing 
\begin{equation}\label{eq:proj}
\tilde{J}_{\boldsymbol{\gamma}_{j+1}} = \|\hat{\mathbf{G}} (\dot{\mathbf{f}}_{{j+1}} - \dot{\mathbf{f}}_{\boldsymbol{\gamma}_{j+1}}^{\text{proj}})\|_2^2,
\end{equation}
where $\dot{\mathbf{f}}_{{j+1}}\in \mathbb{R}^N$ stacks $\dot{{f}}_{{j+1}}(t_k)$ for all samples $t_k,\ k\in\{1,\ldots,N\}$ and the premultiplication with $\hat{\mathbf{G}}$ reflects the desire to match the projected error to the predicted error rather than the compensation signal itself. The solution to \eqref{eq:proj} is then given by:
\begin{equation}\label{eq:solproj}
\boldsymbol{\gamma}_{j+1} = (\boldsymbol{\Psi}^\top \hat{\mathbf{G}}^\top \hat{\mathbf{G}}\boldsymbol{\Psi})^{-1} \boldsymbol{\Psi}^\top \hat{\mathbf{G}}^\top \hat{\mathbf{G}}\dot{\mathbf{f}}_{{j+1}},
\end{equation}
with $\boldsymbol{\Psi}=[\boldsymbol{\psi}(\alpha(t_1)),\ldots,\boldsymbol{\psi}(\alpha(t_N))]^\top$. 

Update law~\eqref{eq:update} and projection~\eqref{eq:solproj} are the two main components to compensating for $\alpha$-domain disturbances, shown in Algorithm~\ref{algo:alpha}. The next section proves that Algorithm~\ref{algo:alpha} leads to monotonic convergence of the compensation signal.
    \subsection{Convergence}\label{sec:convergence}
    The following lemma states the conditions under which the compensation signal converges monotonically.
    \begin{lemma}\label{th:singular}
        Using update law~\eqref{eq:update} with $L(q)$ and $Q(q)$ causal and stable, followed by projection~\eqref{eq:solproj}, the compensation signal $\dot{f}_{\boldsymbol{\gamma}_j}^{\text{proj}}(\alpha(t_k))$ is monotonically convergent in the two-norm as $j\to \infty$ if and only if \begin{equation}\label{eq:convergence}
            \overline{\sigma}\left(\boldsymbol{\Psi} (\boldsymbol{\Psi}^\top \hat{\mathbf{G}}^\top \hat{\mathbf{G}}\boldsymbol{\Psi})^{-1} \boldsymbol{\Psi}^\top \hat{\mathbf{G}}^\top \hat{\mathbf{G}}\mathbf{Q} (\mathbf{I}-\mathbf{L}\mathbf{G})\right) < 1,
        \end{equation}
        where $\overline{\sigma}$ denotes the maximum singular value and $\mathbf{G}$, $\mathbf{Q}$, and $\mathbf{L}$ are the impulse response matrices of $G(q)$, $Q(q)$, and $L(q)$, respectively, in the form of~\eqref{eq:impulse}. 
    \end{lemma}
    \textit{Proof~\cite{Boeren2016a}:} see~\ref{ap:singular}.$\hfill\blacksquare$
This convergence condition hinges upon the design of $L(q)$ and $Q(q)$. The following theorem provides a frequency-domain convergence condition that facilitates the design of these filters later. 
\begin{theorem}\label{th:freq}
    Using update law~\eqref{eq:update} with $L(q)$ and $Q(q)$ causal and stable, followed by projection~\eqref{eq:solproj}, the compensation signal $\dot{f}_{\boldsymbol{\gamma}_j}^{\text{proj}}(\alpha(t_k))$ is monotonically convergent in the two-norm as $j\to \infty$ if \begin{equation}\label{eq:convergence_freq}
        \sup_{\omega\in[0,\pi]} |Q(e^{j\omega}) (1-L(e^{j\omega})G(e^{j\omega}))| < 1.
    \end{equation}
\end{theorem}
\textit{Proof~\cite{Boeren2016a}:} See~\ref{app:th}.$\hfill\blacksquare$\\
In contrast to~\cite{Boeren2016a}, which addresses convergence of compensation functions for single-input single-output (SISO) systems in a closed loop setting, Lemma~\ref{th:singular} and Theorem~\ref{th:freq} apply to the open-loop feedforward scheme depicted in Figure~\ref{fig:control_scheme}, where the two shear references are simultaneously updated by a single compensation function through~\eqref{eq:ff}.

The next section provides some guidelines for designing filters $L(q)$ and $Q(q)$ that satisfy Theorem~\ref{th:freq}, among other implementation aspects. Moreover, Algorithm~\ref{algo:unified} summarizes the unified framework of feedforward control for piezo-stepper actuators, combining the compensation of $\alpha$-domain disturbances with rate-dependent hysteresis compensation.

\begin{algorithm}[tb]
    \caption{Unified framework: flexible feedforward control of piezo-stepper actuators}\label{algo:unified}
    \begin{algorithmic}[1]
    \Require Initial waveforms~\eqref{eq:waveforms}, constant hysteresis models~\eqref{eq:hystconst}, range $\mathcal{F}$ of application-relevant drive frequencies~\eqref{eq:Fgrid}.
    \State Perform Algorithm~\ref{algo:datacol} to collect datasets $\mathcal{D}_e$.
    \State Obtain hysteresis models $\hat{M}_{\boldsymbol{\theta}_{e^\iota}}$ from~\eqref{eq:hysteresis_fit}.% and~\eqref{eq:hyst_model}.
    \State Solve~\eqref{eq:opt} for each $f_\alpha\in\mathcal{F}$ to obtain $g_{e,\max}^{\text{LUT}}(f_\alpha)$ and $g_{e,\min}^{\text{LUT}}(f_\alpha)$ in~\eqref{eq:g}.
    \State Obtain a model $\hat{G}\approx G$ using~\eqref{eq:average_G}.
    \State Perform Algorithm~\ref{algo:alpha} in both directions to obtain modified shear references $\dot{\tilde{r}}_{S_i}(\alpha)$ in~\eqref{eq:ff_dirdep}.% for modified references~\eqref{eq:ff}. 
    \State \Return Control law~\eqref{eq:controllaw2} with models $\hat{M}_{\boldsymbol{\theta}_{e^\iota}}$ and modified shear references $\dot{\tilde{r}}_{S_i}(\alpha)$, with~\eqref{eq:waveforms} and~\eqref{eq:g}.
        \end{algorithmic}
    \end{algorithm}
\subsection{Implementation aspects}\label{sec:impl_alpha}
This section discusses practical considerations for the implementation of Algorithm~\ref{algo:alpha}, starting with obtaining $\hat{G}(q)$.
\subsubsection{Identification of $\hat{G}(q)\approx G(q)$}\label{sec:G}
To obtain a model $\hat{G}(q)\approx G(q)$, which is a prerequisite for Algorithm~\ref{algo:alpha}, the first step is to measure a frequency response function $G(e^{j\omega}):=y(e^{j\omega})/y_\text{true}(e^{j\omega})$. This cannot be done directly as $y_\text{true}$ is unknown. Instead, an indirect approach is taken, where $G_{S_1}$ and $G_{S_2}$ are measured, defined as \begin{equation}
G_{S_i}(e^{j\omega}) := \frac{y(e^{j\omega})}{u_{S_i}(e^{j\omega})}.
\end{equation}
These systems are measured using standard open-loop system identification, where $u_{S_i}$ is excited by a random-phase multisine signal while in contact with the mover, and the mover position $y$ is recorded, in a separate experiment per shear. The data is averaged out over multiple realizations of the random-phase multisines to obtain Best Linear Approximations (BLAs) $\hat{G}_{S_1,\text{BLA}}(e^{j\omega})$ and $G_{S_2,\text{BLA}}(e^{j\omega})$, see~\cite{Pintelon2012} for details.
The frequency response function of the sensor is then approximated as the scaled average \begin{equation}\label{eq:average_G}
    \hat{G}(e^{j\omega}) = \frac{T_s}{e^{j\omega}-1} \cdot \frac{1}{2 c_G}  \left(\hat{G}_{S_1,\text{BLA}}(e^{j\omega})+\hat{G}_{S_2,\text{BLA}}(e^{j\omega})\right),
\end{equation}
where the first term represents a discrete-time integrator and $c_G$ is a scaling factor defined as \begin{equation}
    c_G = \frac{1}{2} \left(\hat{G}_{S_1,\text{BLA}}(e^{j \omega_0})+\hat{G}_{S_2,\text{BLA}}(e^{j\omega_0})\right),
\end{equation}
with $\omega_0$ the lowest measured frequency. This approximation assumes that the sensor $G(q)$ behaves as a discrete-time integrator at low frequencies. Section~\ref{sec:discussion} discusses the limitations of this approximation. Finally, a low-order parametric fit $\hat{G}(q)$ is made of $\hat{G}(e^{j\omega})$.
\subsubsection{Design of $L(q)$ and $Q(q)$}\label{sec:lti_design}
The learning filter $L({q})$ and the robustness filter $Q(q)$ must satisfy condition~\eqref{eq:convergence_freq} and be causal and stable. First, define $L(q)= \beta q^{-d} \hat{G}^{-1}(q)$, where $\beta\in(0,1]$ is a learning gain and $d$ is the relative degree of $\hat{G}(q)$. Lower values of $\beta$ lead to better mitigation of trial-varying disturbances at the cost of slower convergence, see~\cite{OOMEN2017134}. Secondly, design $Q(q)$ such that condition~\eqref{eq:convergence_freq} holds, e.g., by parametrizing $Q(q)$ as a lowpass filter.

\subsubsection{Direction-dependent compensation functions}
Mechanical misalignments lead to a disturbance that is repeatable in the $\alpha$-domain, but is dependent on the direction of motion. Indeed, when the clamps push the shears onto the mover at an angle, this results in a different force depending on whether the shear is expanding or contracting. Therefore, Algorithm~\ref{algo:alpha} is performed twice: once using a positive drive frequency $f_\alpha$ and once using $-f_\alpha$. This results in two compensation functions, $f_{\boldsymbol{\gamma},+}^{\text{proj}}$ and $f_{\boldsymbol{\gamma},-}^{\text{proj}}$. The reference for control law~\eqref{eq:controllaw2} then becomes \begin{equation}\label{eq:ff_dirdep}
        \dot{\tilde{r}}_{S_i}(\alpha) = \begin{cases}
        \dot{r}_{S_i}(\alpha) + \dot{f}_{\boldsymbol{\gamma},+}^{\text{proj}}(\alpha) & \text{if } f_\alpha \geq 0,\\
        \dot{r}_{S_i}(\alpha) + \dot{f}_{\boldsymbol{\gamma},-}^{\text{proj}}(\alpha) & \text{if } f_\alpha < 0.
        \end{cases}
    \end{equation}

\section{Experimental results}\label{sec:experimental}
This section shows experimentally that the developed approach achieves high positioning performance for piezo-stepper actuators despite hysteresis and mechanical misalignments. The experimental setup is described in Section~\ref{sec:setup}.

First, the implementation of the hysteresis compensation method of Section~\ref{sec:hysteresis} is explained, before the compensation of mechanical misalignments of Section~\ref{sec:misalignments} is addressed. 
\subsection{Hysteresis compensation}\label{sec:hysteresis_exp}
\begin{figure}
    \centering
    \includegraphics[width=\linewidth]{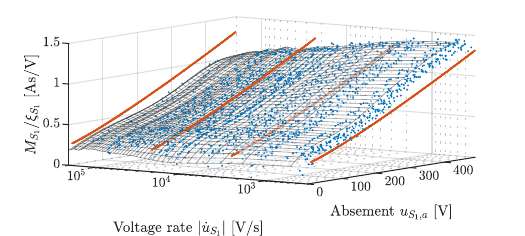}
    \caption{Measurements of the hysteresis function for the first shear (\protect\bluedotfill), with the fitted rate-dependent model $\hat{M}_{S_1,\text{LUT}}$ (\protect\whiterectangle). A rate-independent Ramberg-Osgood model~\cite{Strijbosch2023} (\protect\redline), fitted on all data and plotted along four voltage rates, does not accurately reflect the measurements at high or low voltage rates, whereas the developed rate-dependent model matches the recorded data quite well. The data, sub-sampled for visibility, results from a single experiment described by Algorithm~\ref{algo:datacol} with $F=52$ different drive frequencies between 0.3 Hz and 100 Hz in either direction.} \label{fig:hysteresis_model}
    \end{figure}

Following the procedure in Section~\ref{sec:hysteresis}, hysteresis models $\hat{M}_{{e},\text{LUT}}$ are created measurements of the currents. A total of $F=52$ drive frequencies between 0.3 and 100 Hz are chosen. This choice for $F$ is based on empirical assessment of the smoothness of the resulting hysteresis data in the $|\dot{u}_e|$-direction, see Figure~\ref{fig:hysteresis_model}. This figure displays the resulting data and the fitted model $\hat{M}_{S_1,\text{LUT}}$ for shear $S_1$, demonstrating a very good match between the data and this model for all different voltage rates and absements. For comparison, the rate-independent Ramberg-Osgood model presented in~\cite{Strijbosch2023} is also shown, drawn in red for several voltage rates. This model is given by \begin{equation}
\hat{M}_{S_1,\text{RO}}(u_{S_1,a},\mathbf{h}) = h_1+h_2 u_{S_1,a}^{h_3},
\end{equation}
with parameters $\mathbf{h}=[h_1,h_2,h_3]^\top$ given by the unique solution to the separable least-squares problem \begin{equation}
    \mathbf{h} =\arg\min_{\mathbf{h}} \sum_{i=1}^n \left(\hat{M}_{S_1,\text{RO}}(u_{S_1,a,i},\mathbf{h})-m_{S_1,i} \right)^2,
\end{equation}
see~\cite{Strijbosch2023} for details. Since the Ramberg-Osgood model is rate-independent, it does not accurately reflect the hysteresis behavior at low or high voltage rates, whereas the rate-dependent model presented in this paper does, as clearly seen in Figure~\ref{fig:hysteresis_model}. 

\begin{figure}
        \centering
        \begin{subfigure}{0.48\linewidth}
            \centering
            \includegraphics[width=\linewidth]{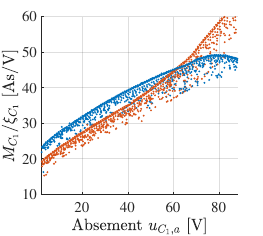}
            \caption{Clamp $C_1$}
        \end{subfigure}
        \hfill
        \begin{subfigure}{0.48\linewidth}
            \centering
            \includegraphics[width=\linewidth]{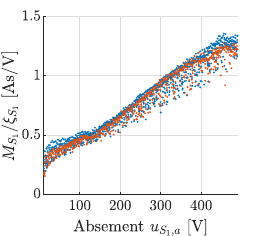}
            \caption{Shear $S_1$}\label{fig:direction_dependent_b}
        \end{subfigure}
        \caption{Hysteresis measurements of forwards (\protect\bluedotfill) and backwards (\protect\reddotfill) motions for all measured voltage rates together. (a) The first clamp shows much more significant direction-dependency than (b) the first shear. Note that (b) is a side-view of the data in Figure~\ref{fig:hysteresis_model}.}\label{fig:direction_dependent}
    \end{figure}

The hysteresis measurements of the first clamp and the first shear are displayed in Figure~\ref{fig:direction_dependent}, which shows a clear direction-dependency in the clamp. Similar direction-dependent hysteresis behaviour is seen in the second clamp, but only to a limited extent in either of the shears. The direction-dependency is accounted for by separately modeling $\hat{M}_{\boldsymbol{\theta}_{e^{+}}}$ and $\hat{M}_{\boldsymbol{\theta}_{e^{-}}}$, see~\eqref{eq:M_dirdep} and~\eqref{eq:hyst_model}.
% It is hypothesized that this {\color{red} todo:why?}. 
% Therefore, separate hysteresis models are created for elements $\overline{e}\in\overline{\Omega}\subset \tilde{\Omega}$, with \begin{equation}
% \overline{\Omega} = \left\{S_1,S_2,C_1^{(-)},C_2^{(-)},C_1^{(+)},C_2^{(+)}\right\}.
% \end{equation}

Finally, these models are used to create lookup tables~\eqref{eq:g} as explained in Section~\ref{sec:drift} to mitigate integrator drift. From these tables, it follows that the reference stroke required to reach the voltage bounds gradually decreases from \SI{3.8}{\micro\meter} to \SI{3}{\micro\meter} between drive frequencies of 1 and 100 Hz. With these elements in place, control law~\eqref{eq:controllaw2} allows compensation of hysteresis in the piezo-stepper actuator.
    
    \subsection{Compensation of $\alpha$-dependent disturbances}
\begin{figure}
        \centering
        \includegraphics[width=\linewidth]{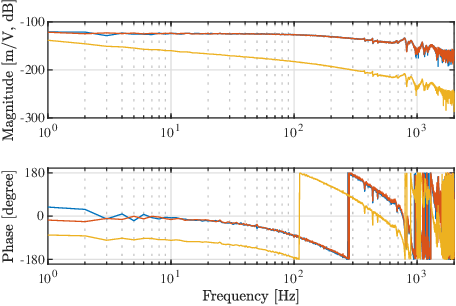}
        \caption{Bode plots of the shear BLAs $\hat{G}_{S_1,\text{BLA}}(e^{j\omega})$ (\protect\blueline) and $\hat{G}_{S_2,\text{BLA}}(e^{j\omega})$ (\protect\redline), used to approximate the scaled sensor dynamics $c_G \hat{G}(e^{j\omega})$ (\protect\yelline).}\label{fig:frfs}
        \end{figure}

\begin{figure}
            \centering
            \includegraphics[width=\linewidth]{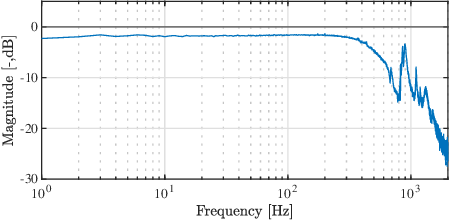}
            \caption{Magnitude plot of $Q(e^{j\omega}) (1-L(e^{j\omega})\hat{G}(e^{j\omega}))$. The condition for convergence of ILC in Theorem~\ref{th:freq} holds since this magnitude is below 0 dB for all frequencies.}\label{fig:ilc_condition}
            \end{figure}

    With hysteresis compensated through control law~\eqref{eq:controllaw2}, the next step is compensating $\alpha$-dependent disturbances. First, the LTI model $\hat{G}(q)$ is identified as described in Section~\ref{sec:G}, see Figure~\ref{fig:frfs}. Subsequently, $L(q)$ and $Q(q)$ are designed as described in Section~\ref{sec:lti_design}, with $\beta=0.2$ and $Q(q)$ a 2nd order lowpass filter with 500 Hz cutoff frequency. Finally, Algorithm~\ref{algo:alpha} is followed with drive frequency $f_\alpha=2$ Hz. The convergence condition from Theorem~\ref{th:freq} is visualized in Figure~\ref{fig:ilc_condition}, showing that the designed $Q$ and $L$ filter together lead to a convergent ILC scheme. The next section presents the results.
\subsection{Results}
This section compares the performance of three different feedforward control strategies: \begin{enumerate}[label=S\arabic*:]
    \item Traditional feedforward control: control law~\eqref{eq:controllaw2} with reference displacements $\rho_e(\alpha,f_\alpha)$ and constant hysteresis models $\hat{M}_e=c_{M,e}$.
    \item Rate-dependent hysteresis compensation: control law~\eqref{eq:controllaw2} with reference displacements $\rho_e(\alpha,f_\alpha)$ and rate-dependent hysteresis models~\eqref{eq:hyst_model}.
    \item Rate-dependent hysteresis compensation in conjunction with $\alpha$-domain disturbance compensation: control law~\eqref{eq:controllaw2} with clamp reference displacements $\rho_{C_i}(\alpha,f_\alpha)$, and modified shear reference displacements $\dot{\tilde{r}}_{S_i}$ from Algorithm~\ref{algo:alpha}, and rate-dependent hysteresis models~\eqref{eq:hyst_model}.
\end{enumerate}
The performance of traditional strategy S1 is shown in Figure~\ref{fig:alpha-dependent} and discussed in Section~\ref{sec:alpha-dependent}. The following section presents the results of strategies S2 and S3 for a single drive frequency $f_\alpha=2$ Hz, after which all three strategies are compared for a range of drive frequencies.% in Section~\ref{sec:performance_compare}.

\subsubsection{Performance at $f_\alpha=2$ Hz}
\begin{figure}
    \centering
    \includegraphics[width=\linewidth]{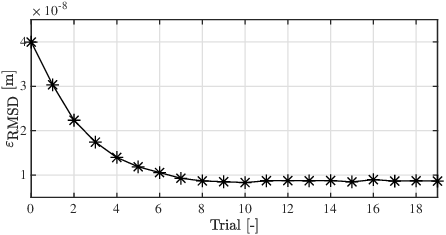}
    \caption{Convergence of the root-mean-square deviation of the tracking error during Algorithm~\ref{algo:alpha}, over the course of twenty trials. Note that hysteresis compensation is active in all trials through control law~\eqref{eq:controllaw2}.
    }\label{fig:ilc_converge}
    \end{figure}

\begin{figure}
        \centering
        \includegraphics[width=\linewidth]{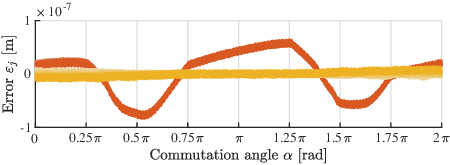}
        \caption{Experimentally measured tracking error of the first (\protect\redline, S2) and twentieth (\protect\yelline, S3) iteration of Algorithm~\ref{algo:alpha}, with hysteresis compensation active in both experiments. The six forwards steps are displayed separately, each with different color saturation and their mean set to zero.}\label{fig:error_trials}
        \end{figure}

% Figure~\ref{fig:ilc_converge} illustrates the reduction in root mean square deviation (RMSD) of the tracking error over twenty iterations of Algorithm~\ref{algo:alpha}. The first trial corresponds to control strategy S1 and the last corresponds to strategy S3. The error decreases from \SI{40}{\nano\meter} to \SI{8.7}{\nano\meter}. The tracking error of the first and last iterations is compared in Figure~\ref{fig:error_trials}, showing that the learned compensation function successfully eliminates the disturbance. The remaining error largely consists of an oscillation at approximately 3250 Hz, as seen in the reverse cumulative amplitude spectrum in Figure~\ref{fig:piezo_spectrum}, possibly caused by flexible dynamics. Since this oscillation does not repeat in the $\alpha$-domain, it remains uncorrected by ILC, contributing approximately \SI{3.5}{\nano\meter} to the RMSD error. Frequencies below $f_\alpha$ exceed the $\alpha$-domain and are also not targeted by the compensation approach. Overall, the significant reduction in tracking error demonstrates the effectiveness of the developed method in achieving highly accurate positioning for piezo-stepper actuators. The next section demonstrates that these results generalize to other drive frequencies as well.
Figure~\ref{fig:ilc_converge} shows the reduction in root-mean-square deviation (RMSD) of the tracking error over twenty iterations of Algorithm~\ref{algo:alpha}, showing a reduction from \SI{40}{\nano\meter} to \SI{8.7}{\nano\meter} over the iterations. The first and last iterations correspond to control strategies S2 and S3, respectively. Figure~\ref{fig:error_trials} compares their tracking errors, confirming that the learned compensation function effectively eliminates the disturbance. 

\begin{figure}
                \centering
                \includegraphics[width=\linewidth]{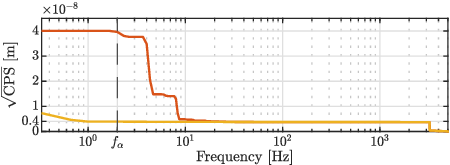}
                \caption{Reverse cumulative amplitude spectrum of the tracking error $\varepsilon$ in the first (\protect\redline) and last (\protect\yelline) iteration of Algorithm~\ref{algo:alpha} at $f_\alpha=2$ Hz. As most of the remaining error appears below the drive frequency (\protect\blackdash), which exceeds the $\alpha$-domain, or at very high frequencies, the $\alpha$-domain disturbances are considered eliminated.}\label{fig:piezo_spectrum}
                \end{figure}

The remaining error is dominated by an oscillation around 3250 Hz, as seen in the reverse cumulative amplitude spectrum in Figure~\ref{fig:piezo_spectrum}, likely due to flexible dynamics. Since this oscillation does not repeat in the $\alpha$-domain, it remains uncorrected by ILC, contributing approximately \SI{3.5}{\nano\meter} to the RMSD error. Additionally, frequencies below $f_\alpha$ exceed the $\alpha$-domain and are not addressed by the compensation. The next section shows that these performance improvements generalize to other drive frequencies as well.


\subsubsection{Performance across different drive frequencies}\label{sec:performance_compare}
\begin{figure}
            \centering
            \begin{subfigure}{0.48\linewidth}
                \centering
                \includegraphics[width=\linewidth]{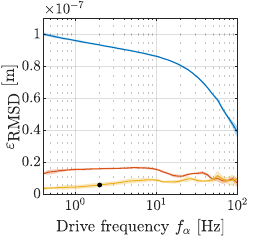}
                \caption{Forwards stepping.}
            \end{subfigure}
            \hfill
            \begin{subfigure}{0.48\linewidth}
                \centering
                \includegraphics[width=\linewidth]{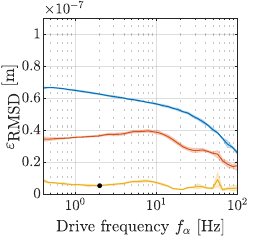}
                \caption{Backwards stepping.}
            \end{subfigure}
            \caption{Root-mean-square deviation of the position error, averaged over three subsequent steps, for different drive frequencies. Traditional feedforward control (\protect\blueline, S1) leads to consistently large error. Hysteresis compensation (\protect\redline, S2) by itself results in a performance increase, but more importantly, it is a prerequisite for Algorithm~\ref{algo:alpha} (\protect\yelline, S3), which reduces the error for all drive frequencies after converging solely with $f_{\alpha}=2$ Hz (\protect\blackdotfill). Shaded areas reflect two standard deviations.}\label{fig:performance_compare}
            \end{figure}

\begin{figure}
                \centering
                \begin{subfigure}{0.48\linewidth}
                    \centering
            \includegraphics[width=\linewidth]{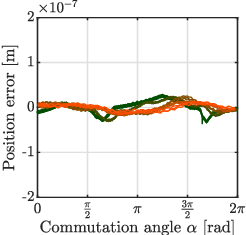}
                    \caption{Strategy S2, forwards stepping.}
                \end{subfigure}
                \hfill
                \begin{subfigure}{0.48\linewidth}
                    \centering
                    \includegraphics[width=\linewidth]{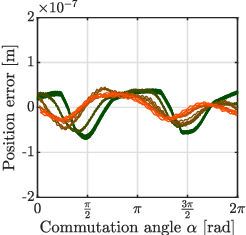}
                    \caption{Strategy S2, backwards stepping.}
                \end{subfigure}
                \begin{subfigure}{0.48\linewidth}
                    \centering
            \includegraphics[width=\linewidth]{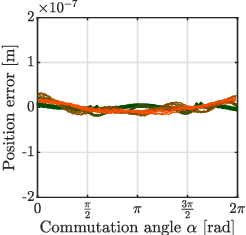}
                    \caption{Strategy S3, forwards stepping.}
                \end{subfigure}
                \hfill
                \begin{subfigure}{0.48\linewidth}
                    \centering
                    \includegraphics[width=\linewidth]{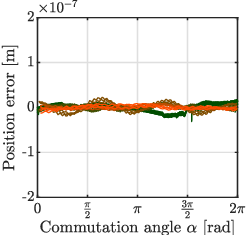}
                    \caption{Strategy S3, backwards stepping.}
                \end{subfigure}
                \caption{Position error of the mover against the commutation angle $\alpha$, after hysteresis compensation (S2, top) and after hysteresis compensation in conjunction with compensation of mechanical misalignments (S3, bottom). The data reflects a range of constant drive frequencies between 0.4 Hz (\protect\dgreenline) and 100 Hz (\protect\dredline), showing three steps per frequency. These plots complement the results of the uncompensated situation S1 in Figure~\ref{fig:alpha-dependent}, showing that the developed approach effectively reduces the tracking error for all drive frequencies.}\label{fig:alpha-dependent_result}
                        \end{figure}

To assess the flexibility of the approach, the performance is compared for drive frequencies ranging from 0.4 to 100 Hz, see Figure~\ref{fig:performance_compare}. When using traditional feedforward strategy S1, the error remains high across all frequencies, see also Figure~\ref{fig:alpha-dependent}. Although rate-dependent hysteresis compensation (S2) improves the performance according to Figure~\ref{fig:performance_compare}, it does not address the $\alpha$-dependent disturbances. Utilizing the compensation function $\dot{f}^{\text{proj}}(\alpha)$ obtained from Algorithm~\ref{algo:alpha} (S3) yields a significant improvement for all drive frequencies: a fifteenfold improvement at $f_\alpha=2$ Hz, when compared to S1. These conclusions are supported by Figure~\ref{fig:alpha-dependent_result}, showing the same data in the $\alpha$-domain for a smaller range of drive frequencies.

Moreover, an interesting difference between forwards and backwards stepping is observed in Figure~\ref{fig:performance_compare}. With traditional feedforward, the performance is better in the backwards direction, but when only hysteresis is compensated, the performance is better in the forwards direction. This suggests that the effect of $\alpha$-dependent disturbances is higher in the backwards direction, but that hysteresis is less pronounced in this direction. The compensation of $\alpha$-dependent disturbances in conjunction with hysteresis compensation (S3) leads to a similar performance in both directions, showing that the developed approach effectively removes both direction-dependency and rate-dependency in the system.
    
                        \subsection{Discussion}\label{sec:discussion}
                        This section addresses several further observations drawn from the experimental results. 
                        % \subsubsection{Impact of drive frequency on performance}
                        % The error peak around 60 Hz in Figure~\ref{fig:performance_compare} is likely attributable to imperfect electromagnetic shielding of the capacitive position sensor. Moreover, the performance increase evident in the blue line of Figure~\ref{fig:performance_compare} for increasing drive frequencies is consistent with the decreased hysteresis measurements $M_{S_1}$ shown in Figure~\ref{fig:hysteresis_model} at higher voltage rates. This is expected, since higher drive frequencies correspond to higher voltage rates when the control law~\eqref{eq:controllaw} is implemented with an imperfect, constant hysteresis model $\hat{M}_{S_1,\text{const}}$ from~\eqref{eq:hystconst} and the reference displacements~\eqref{eq:waveforms}. Additionally, the second-order lowpass filters in both the amplifier (cutoff around 400 Hz) and the capacitive sensor (cutoff around 100 Hz) influence the measured RMSD position error at high drive frequencies. {\color{red} todo:omgooien}
                        
                        \subsubsection{Limitations of the capacitive position sensor}
                        The capacitive sensor $G(q)$ has a cutoff frequency of approximately 100 Hz, significantly impacting the phase and magnitude of the measured mover position at higher drive frequencies. This sensor is used in Algorithm~\ref{algo:alpha} and to assess performance. In Algorithm~\ref{algo:alpha}, the measured mover position is filtered with $L\approx G^{-1}$ before projecting the compensation signal to the $\alpha$-domain, essentially correcting for these sensor dynamics. 
                        
                        However, the sensor dynamics are not corrected for in the performance assessment, leading to a discrepancy between the measured and actual performance. The result is that Figures~\ref{fig:alpha-dependent} and Figure~\ref{fig:alpha-dependent_result} display more variation over frequencies than what can be explained by hysteresis only. Similarly, Figure~\ref{fig:performance_compare} shows a reduced RMSD error for high drive frequencies when using traditional feedforward. This is explained by the sensor dynamics as well, since lowpass effects in the sensor reduce the measured error ripples at high drive frequencies. 

                        \subsubsection{Artifacts in the identified sensor model}
                        The identified sensor model in Figure~\ref{fig:frfs} suggests the presence of flexible dynamics at high frequencies. These dynamics do not reflect the sensor itself but are rather an artifact from the approximation~\eqref{eq:average_G}. By constructing $\hat{G}(q)$ from the BLAs of the shears where voltage is the input, flexible dynamics of the piezo element may end up in $\hat{G}(q)$. This does not pose a problem for the approach: as long as the magnitude of $1 - Q(e^{j\omega})(1 - L(e^{j\omega})G(e^{j\omega}))$ in Figure~\ref{fig:ilc_condition} remains robustly below 0 dB, the ILC scheme converges.
                        
                        \subsubsection{Systematic effects in the hysteresis model}
                        Finally, the identified hysteresis models \(\hat{M}_e(\dot{u}_e,u_{e,a})\) are obtained from measurements where the hysteresis of the piezo elements is unavoidably coupled with systematic contributions from contact dynamics and linear dynamics. Such systematic effects, impacting the piezo displacements while being dependent on voltage rate or absement, end up in \(\hat{M}_e(\dot{u}_e,u_{e,a})\) and are also compensated for by the inversion in control law~\eqref{eq:controllaw2}. This improves performance, and any residual, unmodeled $\alpha$-dependent effects are subsequently targeted by Algorithm~\ref{algo:alpha}. Therefore, although \(\hat{M}_e\) captures more than just hysteresis, the total framework robustly compensates for all these effects, ensuring accurate positioning for arbitrary drive frequencies.

\section{Conclusion}\label{sec:conclusion}
The developed control framework for piezo-stepper actuators achieves accurate and flexible positioning by compensating hysteresis and misalignments. The framework is validated on one specific type of piezo-stepper actuator. It is directly applicable to other piezo and piezo-stepper designs that suffer from hysteresis or mechanical misalignments. First, a rate-dependent hysteresis function is modeled from data and used in a feedforward control law that decouples the piezo input signals from their history. Next, iterative learning control (ILC) is applied to learn a compensation function added to the shear waveforms, ensuring a constant mover velocity even when the piezo elements are imperfectly aligned.

The approach relies on lookup tables and a low number of arithmetic operations during real-time evaluation, directly enabling implementation on embedded control platforms.  Moreover, as this is a feedforward approach, its improvements in positioning accuracy enable the reduction of the feedback gain in closed loop piezo-stepper control, reducing the amplification of measurement noise. In some cases, this may allow for cost savings on high-precision position sensors or even eliminate the need for a position sensor entirely.

Finally, experimental results confirm that the improvement in positioning accuracy is robust to changes in the reference velocities, further increasing the industrial applicability of the method. These results show that the developed feedforward approach achieves accurate positioning of piezo-stepper actuators, enabling their use in applications that require high precision, stiffness, and task flexibility, without expensive position sensors.

% \section{Acknowledgements}
% The authors gratefully acknowledge the contributions to this paper through a challenge-based learning project by Leontine Aarnoudse, Lennart Blanken, Koen Classens, Max van Haren, Johan Kon, Paul Tacx, Rodrigo A. González, Mathyn van Dael, Karlijn van Acht, Sara Betancur Giraldo, Mees Bieling, Lowe Blom, Timo Corvers, Marjolein Daanen, Stern Eichperger, Rainier Heijne, Maarten van der Hulst, Adis Husanović, Tim Jansen, Tim de Keijzer, Andrey Kharitenko, Marijn van Noije, Radu Pîrvan, Finn Ruijters, Rikuto Suzuki, Matthijs Turk, Frank Vlaar, Luuk van Vliet, Michiel Wind, Guido Wolfs, and Pieter van Wonderen.

\section*{CRediT authorship contribution statement}
\textbf{Max van Meer}: Conceptualization, Methodology, Validation, Formal Analysis, Writing - Original Draft - Review \& Editing, Visualization. \textbf{Tim van Meijel}: Conceptualization, Methodology, Software, Validation, Formal Analysis, Investigation, Visualization. \textbf{Emile van Halsema}: Resources, Writing - Review \& Editing, Supervision. \textbf{Edwin Verschueren}: Resources, Writing - Review \& Editing, Supervision. \textbf{Gert Witvoet}: Writing - Review \& Editing, Supervision. \textbf{Tom Oomen}: Writing - Review \& Editing, Supervision.
\appendix
\section{Proof of Lemma~\ref{th:singular}}\label{ap:singular}
First, the projection~\eqref{eq:solproj} is substituted in $\dot{\mathbf{f}}_{\boldsymbol{\gamma}_{j+1}}^{\text{proj}}=\boldsymbol{\Psi}\boldsymbol{\gamma}_{j+1}$ to obtain \begin{equation}
    \dot{\mathbf{f}}_{\boldsymbol{\gamma}_{j+1}}^{\text{proj}}=\boldsymbol{\Psi} (\boldsymbol{\Psi}^\top \hat{\mathbf{G}}^\top \hat{\mathbf{G}}\boldsymbol{\Psi})^{-1} \boldsymbol{\Psi}^\top \hat{\mathbf{G}}^\top \hat{\mathbf{G}}\dot{\mathbf{f}}_{{j+1}}.
\end{equation}
With update law~\eqref{eq:update}, this becomes \begin{equation}\label{eq:proof1_tmp}
    \dot{\mathbf{f}}_{\boldsymbol{\gamma}_{j+1}}^{\text{proj}}=\boldsymbol{\Psi} (\boldsymbol{\Psi}^\top \hat{\mathbf{G}}^\top \hat{\mathbf{G}}\boldsymbol{\Psi})^{-1} \boldsymbol{\Psi}^\top \hat{\mathbf{G}}^\top \hat{\mathbf{G}}\mathbf{Q} (\dot{\mathbf{f}}_{j}^{\text{proj}}+\mathbf{L} \boldsymbol{\varepsilon}_j).
\end{equation}
Next, express the error~\eqref{eq:error} as
\begin{equation}
    {\boldsymbol{\varepsilon}}_{j}=  (\hat{\mathbf{G}}-\mathbf{G})\dot{\mathbf{r}}_{S_i} - \mathbf{G}\dot{\mathbf{f}}_{\boldsymbol{\gamma}_j}^{\text{proj}} - \mathbf{d}_{j},
\end{equation}
and substitute it into~\eqref{eq:proof1_tmp} to obtain
\begin{equation}\label{eq:proof1_tmp2}
    \begin{aligned}
        \dot{\mathbf{f}}_{\boldsymbol{\gamma}_{j+1}}^{\text{proj}}=&\boldsymbol{\Psi} (\boldsymbol{\Psi}^\top \hat{\mathbf{G}}^\top \hat{\mathbf{G}}\boldsymbol{\Psi})^{-1} \boldsymbol{\Psi}^\top \hat{\mathbf{G}}^\top \hat{\mathbf{G}}\mathbf{Q} \\
        &\cdot \left((\mathbf{I}-\mathbf{L}\mathbf{G})\dot{\mathbf{f}}_{j}^{\text{proj}}+\mathbf{L} \left((\hat{\mathbf{G}}-\mathbf{G})\dot{\mathbf{r}}_{S_i} - \mathbf{d}_{j}\right)\right).
    \end{aligned}
    \end{equation}
    Define the matrix
    \begin{equation}
        \mathbf{A} = \mathbf{D} \mathbf{Q} \left( \mathbf{I} - \mathbf{L} \mathbf{G} \right),
    \end{equation}
    and the vector
    \begin{equation}
        \mathbf{b}_j = \mathbf{D} \mathbf{Q} \mathbf{L} \left( (\hat{\mathbf{G}} - \mathbf{G}) \dot{\mathbf{r}}_{S_i} - \mathbf{d}_{j} \right), 
    \end{equation}
    where \begin{equation}\label{eq:D}
\mathbf{D}:=\boldsymbol{\Psi} \left( \boldsymbol{\Psi}^\top \hat{\mathbf{G}}^\top \hat{\mathbf{G}} \boldsymbol{\Psi} \right)^{-1} \boldsymbol{\Psi}^\top \hat{\mathbf{G}}^\top \hat{\mathbf{G}}.
    \end{equation}
    Then, the update equation~\eqref{eq:proof1_tmp2} simplifies to
    \begin{equation}
        \dot{\mathbf{f}}_{\boldsymbol{\gamma}_{j+1}}^{\text{proj}} = \mathbf{A} \dot{\mathbf{f}}_{j}^{\text{proj}} + \mathbf{b}_j.
    \end{equation}
    Since $\dot{\mathbf{r}}_{S_i}$, $\mathbf{d}_{j}$, and $(\hat{\mathbf{G}} - \mathbf{G})$ are bounded, the vector $\mathbf{b}_j$ is bounded. Moreover, when $\overline{\sigma}(\mathbf{A}) < 1$, it follows that $\| \mathbf{A}^j \|_2 \to 0$ as $j \to \infty$. By iterating the update equation, this becomes
    \begin{equation}
        \dot{\mathbf{f}}_{j}^{\text{proj}} = \mathbf{A}^j \dot{\mathbf{f}}_{0}^{\text{proj}} + \sum_{k=0}^{j-1} \mathbf{A}^{j-1-k} \mathbf{b}_k.
    \end{equation}
    The term $\mathbf{A}^j \dot{\mathbf{f}}_{0}^{\text{proj}}$ tends to zero because $\| \mathbf{A}^j \|_2 \to 0$. Since $\mathbf{b}_k$ is bounded and $\| \mathbf{A}^{j-1-k} \|_2$ decreases with each iteration, the summation converges to a finite limit. Therefore, when the condition in~\eqref{eq:convergence} holds, the sequence $\| \dot{\mathbf{f}}_{j}^{\text{proj}} \|_2$ decreases monotonically and converges in the two-norm as $j \to \infty$.
$\hfill\blacksquare$

%  DO NOT REMOVE. THE FOLLOWING PROOF ONLY HOLDS IF L IS CAUSAL, SEE BOEREN
\section{Proof of Theorem~\ref{th:freq}}\label{app:th}
The condition~\eqref{eq:convergence} of Lemma~\ref{th:singular} is bounded by 
\begin{equation}\label{eq:convergence_bound}
        \overline{\sigma}\left(\mathbf{D} \mathbf{Q} (\mathbf{I}-\mathbf{L}\mathbf{G})\right)\leq \overline{\sigma}\left(\mathbf{D}\right) \overline{\sigma}\left(\mathbf{Q} (\mathbf{I}-\mathbf{L}\mathbf{G})\right),
    \end{equation}
    with $\mathbf{D}$ defined in~\eqref{eq:D}. Hence, condition~\eqref{eq:convergence} is satisfied if $\overline{\sigma}\left(\mathbf{D}\right) \overline{\sigma}\left(\mathbf{Q} (\mathbf{I}-\mathbf{L}\mathbf{G})\right)\leq 1$. This is the case if~\eqref{eq:convergence_freq} holds and $\overline{\sigma}\left(\mathbf{D}\right)=1$. To see that $\overline{\sigma}\left(\mathbf{D}\right)=1$, note that $\mathbf{D}$ is idempotent, i.e., $\mathbf{D}^2=\mathbf{D}$. Since all eigenvalues $\lambda_i$ of an idempotent matrix are either $0$ or $1$, and $\mathbf{D}$ is not a zero matrix, $\mathbf{D}$ has a largest eigenvalue $\overline{\lambda}=1$, and hence, $\overline{\sigma}(\mathbf{D})=1$. As such, condition~\eqref{eq:convergence} reduces to \begin{equation}
        \overline{\sigma}\left(\mathbf{Q} (\mathbf{I}-\mathbf{L}\mathbf{G})\right)\leq 1.
    \end{equation}
    Finally, it follows directly from~\cite[Theorem 8]{Norrlof2002a} that when $L(q)$ and $Q(q)$ are stable and causal, it holds that \begin{equation}
        \sup_{\omega\in[0,\pi]} |Q(e^{j\omega}) (1-L(e^{j\omega})G(e^{j\omega}))| < 1 \implies  \overline{\sigma}\left(\mathbf{Q} (\mathbf{I}-\mathbf{L}\mathbf{G})\right)\leq 1.
    \end{equation}
Hence,~\eqref{eq:convergence_freq} is a sufficient condition for convergence.$\hfill\blacksquare$

\addtolength{\textheight}{-12cm}  
\bibliographystyle{IEEEtran}
% \bibliographystyle{plain}
%\bibliography{articles_copy,bookspersonal_copy,booksnotpersonal_copy,reports_copy}
\bibliography{library}

\end{document}

%% file: tikz_set.tex
\usepackage{tikz}
\usepackage{tikzscale}

\definecolor{myorange}{cmyk}{0,0.35,0.85,0} 
\definecolor{mypurple}{cmyk}{0.5,1,0,0} 

\definecolor{matblue1}{rgb}{0,0.4470,0.7410}
\definecolor{matred1}{rgb}{0.85,0.325,0.098}
\definecolor{matyel1}{rgb}{0.9290, 0.6940, 0.1250}
\definecolor{matpur1}{rgb}{0.4940, 0.1840, 0.5560}
\definecolor{matgre1}{rgb}{0.4660, 0.6740, 0.1880}
\definecolor{matblue2}{rgb}{0.3010, 0.7450, 0.9330}
\definecolor{matred2}{rgb}{0.6350, 0.0780, 0.1840}
\definecolor{matgrey1}{rgb}{0.5, 0.6, 0.7}
\definecolor{matpink1}{rgb}{1, 0.07, 0.65}
\definecolor{matblue3}{rgb}{0.07, 0.62, 1}
\definecolor{gray09}{rgb}{0.9, 0.9, 0.9}

\definecolor{dgreen}{rgb}{0.1, 0.3, 00}
\definecolor{dred}{rgb}{1, 0.3, 0}

    \definecolor{mblue}{rgb}{0,0.447,0.741}
    \definecolor{mred}{rgb}{0.85,0.325,0.098}
    \definecolor{myellow}{rgb}{0.9290,0.6940,0.1250}
    \definecolor{mmagenta}{rgb}{1,0,1}
    \definecolor{mgreen}{rgb}{0.4460,0.6740,0.1880}
    \definecolor{mgrey}{rgb}{0.6,0.6,0.6}
    \definecolor{mpurple}{rgb}{0.4940, 0.1840, 0.5560}

    \definecolor{matbluel}{rgb}{0,0.6732,1}
    \definecolor{matyeld}{rgb}{0.2787,0.2082,0.0375}

    \tikzset{cross/.style={cross out, draw=black, minimum size=2*(#1-\pgflinewidth), inner sep=0pt, outer sep=0pt}, cross/.default={1pt}}
    
    \usetikzlibrary{shapes}

\newcommand{\blackdash}{\raisebox{2pt}{\tikz{\draw[-,black,dashed,line width = 0.9pt](0,0) -- (3mm,0);}}}

\newcommand{\blueline}{\raisebox{2pt}{\tikz{\draw[-,matblue1,solid,line width = 0.9pt](0,0) -- (3mm,0);}}}
\newcommand{\redline}{\raisebox{2pt}{\tikz{\draw[-,matred1,solid,line width = 0.9pt](0,0) -- (3mm,0);}}}

\newcommand{\bluedotfill}{%
    \raisebox{0.5pt}{%
        \tikz{\fill[matblue1] (0,0) circle (0.5mm);}%
    }%
}
\newcommand{\reddotfill}{%
    \raisebox{0.5pt}{%
        \tikz{\fill[matred1] (0,0) circle (0.5mm);}%
    }%
}
\newcommand{\blackdotfill}{%
    \raisebox{0.5pt}{%
        \tikz{\fill[black] (0,0) circle (0.5mm);}%
    }%
}
\newcommand{\whiterectangle}{%
    \raisebox{0pt}{%
        \tikz{\draw[draw=gray, fill=white, line width=0.5pt] (0,0) rectangle (2mm, 1.5mm);}%
    }%
}

\newcommand{\yelline}{\raisebox{2pt}{\tikz{\draw[-,matyel1,solid,line width = 0.9pt](0,0) -- (3mm,0);}}}
\newcommand{\purline}{\raisebox{2pt}{\tikz{\draw[-,matpur1,solid,line width = 0.9pt](0,0) -- (3mm,0);}}}

\newcommand{\dgreenline}{\raisebox{2pt}{\tikz{\draw[-,dgreen,solid,line width = 0.9pt](0,0) -- (3mm,0);}}}
\newcommand{\dredline}{\raisebox{2pt}{\tikz{\draw[-,dred,solid,line width = 0.9pt](0,0) -- (3mm,0);}}}

\usepackage{tikz}
\usepackage{xcolor}

\newlength{\boxplotlinewidth} 